\documentclass[preprint]{aastex}
\usepackage{epsfig}

\shortauthors{Abbett, Fisher \& Fan} 
\shorttitle{Evolution of Rising, Twisted Magnetic Flux Tubes}

\begin{document}

\title{The Three-dimensional Evolution of Rising, Twisted Magnetic 
   Flux Tubes in a Gravitationally Stratified Model Convection Zone \\}
\author{W. P. Abbett and G. H. Fisher}
\affil{Space Sciences Laboratory, University of California,
    Berkeley, CA 94720-7450}
\and
\author{Y. Fan}
\affil{HAO, National Center for Atmospheric Research, P.O. Box
   3000, Boulder, CO 80307}
\email{yfan@hao.ucar.edu}

\begin{abstract}
We present three-dimensional numerical simulations of 
the rise and fragmentation of twisted, initially horizontal 
magnetic flux tubes which evolve into emerging $\Omega$-loops.  
The flux tubes rise buoyantly through an adiabatically stratified plasma 
that represents the solar convection zone.  The MHD equations are 
solved in the anelastic approximation, and the results are compared 
with studies of flux tube fragmentation in two dimensions.  We find
that if the initial amount of field line twist is below a critical 
value, the degree of fragmentation at the apex of a rising 
$\Omega$-loop depends on its three-dimensional geometry: the greater 
the apex curvature of a given $\Omega$-loop, the lesser the degree of 
fragmentation of the loop as it approaches the photosphere.  Thus, the 
amount of initial twist necessary for the loop to retain its cohesion 
can be reduced substantially from the two-dimensional limit.  The 
simulations also suggest that as a fragmented flux tube emerges 
through a relatively quiet portion of the solar disk, extended 
crescent-shaped magnetic features of opposite polarity should form and 
steadily recede from one another.  These features eventually coalesce 
after the fragmented portion of the $\Omega$-loop emerges through the 
photosphere. 
\end{abstract}


\section{Introduction}

The largest concentrations of magnetic flux on the Sun occur in active
regions.  Great progress has been made over the past decade in understanding
the connections between the magnetic field in active regions, observed at 
the surface of the Sun, to the magnetic field deep in the solar interior.  
Active regions have a bipolar structure, suggesting that they are the tops 
of magnetic flux loops which have risen from deep in the solar interior.  
On average, active regions are oriented in the E-W direction (Hale's 
Polarity Law) suggesting that the underlying field geometry is toroidal.  
The persistence of Hale's law for periods of several years during a given 
solar cycle suggests that magnetic flux must be stored in a relatively 
stable region of the solar interior.  Several stability arguments 
\citep{svb82,vb82,fms93,fms95} show that the only place where such fields 
can be confined stably for periods of several years is below the solar 
convection zone.  On the other hand, if magnetic fields are placed 
any significant distance below the top of the radiative zone, they are 
so stable they could not emerge on the time scale of a solar cycle.  We 
are thus led to the conclusion that the most likely origin of active 
region magnetic fields is from a toroidally oriented field layer residing 
in the ``convective overshoot region'', a thin, slightly convectively 
stable layer just beneath the convection zone.  This layer also seems to 
coincide with the ``tachocline'' \citep{k96,c99}, where the solar rotation 
rate transitions from solid body behavior in the radiative zone, to the 
observed latitudinally dependent rotation rate we see at the Sun's surface.  
This suggests that not only are solar magnetic fields stored in the 
convective overshoot layer, the overshoot layer is also the most likely
site for the solar cycle dynamo \citep{gmd89,dg91,p93,mc97a,mc97b,d97,dc99}.

Over the past decade, most efforts to study the emergence of active region
magnetic fields have employed the ``thin flux tube'' approximation.  This 
model assumes that magnetic flux tubes behave as distinct tube-like entities, 
surrounded by field-free plasma.  The approximation further assumes that the 
tube diameter is small compared to all other length scales in the problem, 
and that pressure balance exists across the tube at all times.  After 
adopting these assumptions, it is straightforward to derive an equation of 
motion for the dynamics of the tube from the momentum equation in MHD.
Thin flux tube models of emerging active regions have proven very successful
in explaining many properties of active regions in terms of flux tube
dynamics in the solar interior.  For example, they have successfully 
explained the variation of active region tilt with respect to the E-W 
direction as a function of solar latitude \citep{dh93,dc93,ffh95},  
the asymmetric orientation of the magnetic field after emergence 
\citep{vp90,msc94,cmv96}, and the observed scatter in tilts as a function of 
active region size \citep{lf96}.

In spite of these successes, recent two-dimensional MHD simulations of 
flux tube emergence have shown results which seem to invalidate many
assumptions that are adopted in the thin flux tube approximation.
\citet{s79} and \citet{lfa96} find that an initially 
buoyant, untwisted flux tube will fragment into two counter-rotating tube
elements which then separate from one another, essentially destroying
the tube's initial identity.  \citet{mie96} and \citet{fzl98} have 
demonstrated via two-dimensional MHD simulations that in order to prevent 
a flux tube from fragmenting, enough twist must be introduced into the 
tube to provide a cohesive force to balance the hydrodynamic forces acting 
to rip it apart.  That critical twist is defined, roughly, by that necessary 
to make the Alfv\'en speed from the azimuthal component of the field at 
least as great as the relative velocity between the tube and the field 
free plasma surrounding it \citep{llf96,em98,fzl98}. 

But when global levels of twist in active regions \citep{pcm95,lfp98} are 
determined from vector magnetograms, the amplitude of the observed twist is 
typically far smaller than this critical value \citep{llp99}.  When plotted 
as a function of active region latitude, the twists exhibit large scatter, 
but superimposed on this apparently random behavior there is a slight, but 
clearly discernible trend for active regions in the northern hemisphere to 
be negatively twisted, while those in the south are positively twisted.

\citet{lfp98} have developed a theoretical model which not only explains the 
latitudinal variation of twist, but also can account for the large 
fluctuations in twist from active region to active region.  In this
model, an initially untwisted flux tube rises through the convection zone
in accordance with the thin flux tube approximation.  Coriolis forces 
acting on convective eddies produce a non-zero average kinetic helicity, 
which is proportional to latitude.  The kinetic helicity acts to ``writhe'' 
the flux tube, which is then twisted in the opposite direction to preserve 
its magnetic helicity.  \citet{lfp98} showed that this model can explain the 
observed data.  Yet the model assumes from the beginning that the thin
flux tube approximation can be used, even for an initially untwisted tube, 
while the two-dimensional  MHD simulations suggest that this is invalid.  
Is there some way out of this quandary, which we dub ``Longcope's Paradox''?

In this paper, we describe MHD simulations of flux tube fragmentation
in three dimensions.  The result of these simulations is that the critical 
degree of twist necessary to prevent fragmentation is reduced dramatically 
by the presence of flux tube curvature, as will be present in an emerging 
$\Omega$-loop.  We find that for a fixed amount of twist, the degree of 
fragmentation is a function of the tube's curvature, and transitions 
asymptotically to the two-dimensional limit as the curvature approaches 
zero.  Even for flux tubes with little to no initial twist, the fragmenting 
magnetic morphology at the apex of an emerging $\Omega$-loop is considerably 
less dispersed than the two-dimensional simulations would indicate --- a 
finding consistent with the simulation results of \citet{dn98}.  It is not 
clear at this time whether our results resolve Longcope's paradox or not, 
but they certainly ameliorate the problem a great deal.

The remainder of this paper is organized as follows:
In Section~\ref{method} we briefly discuss the formalism of the
anelastic approximation employed in our models, together with the
numerical methods used to solve the system of equations.  We also 
describe the range of initial configurations that we use to
explore the relationship between flux tube fragmentation, tube 
geometry, and the initial twist of the field lines.  At the beginning of 
Section~\ref{results}, we define what is meant by a flux tube in 
the context of our three-dimensional MHD simulations, and further 
define a quantitative measure of the degree of fragmentation of 
such a tube.  We then present the results of our numerical 
simulations and discuss the implications of the models.  Finally,
in Section~\ref{conclusions} we summarize our conclusions.

\clearpage

\section{Method}\label{method}

We solve the three-dimensional MHD equations in the anelastic 
approximation \citep{op62,g69} using a portable, 
modularized version of the code of \citet{fzl99}.  The anelastic 
equations result from a scaled-variable expansion of the equations 
of compressible MHD (for details see \citealt{lf99} and references 
therein), and describe variations of a plasma about a stratified, 
isentropic reference state.  This approach is valid for low acoustic 
Mach number plasma below the photosphere ($M \equiv v/c_\mathrm{s} \ll 1$), 
where the Alfv\'en speed $v_\mathrm{a}$ is much less than the local 
sound speed $c_\mathrm{s}$ \citep{g84}.  The primary computational 
advantage of this technique is that fast-moving acoustic waves are 
effectively filtered out of the simulations.  This allows for much 
larger timesteps than would be possible in fully compressible MHD, 
and unlike a Bousinnesq treatment, a non-trivial background stratification 
can be included in the models.  This method is well suited to our 
investigation of the evolution and fragmentation of magnetic flux tubes, 
since our region of interest lies well within the limits of validity of 
this approximation, and since we require many simulations to fully 
explore the relevant parameter space.  It is important to note, however, 
that where $M$ is not $\ll 1$, a fully compressible treatment 
(such as that of \citealt{bsn99}) is required.  
The equations of anelastic MHD are as follows:
\begin{equation}
\nabla\cdot\left(\rho_0\mathbf{v}\right)=0
\end{equation}
\begin{equation}
\rho_0\left(\frac{\partial\mathbf{v}}{\partial{t}}+
   \mathbf{v}\cdot\nabla\mathbf{v}\right)=-\nabla p_1+\rho_1\mathbf{g}+ 
   \frac{1}{4\pi}\left(\nabla\times\mathbf{B}\right)\times\mathbf{B}+ 
   \nabla\cdot\mathbf{\Pi} 
\end{equation}
\begin{equation}
\rho_0 T_0\left(\frac{\partial s_1}{\partial{t}}+\mathbf{v}\cdot\nabla
   \left(s_0+s_1\right)\right)=\nabla\cdot\left(K\rho_0 T_0\nabla s_1
   \right)+\frac{\eta}{4\pi}\left|\nabla\times\mathbf{B}\right|^2+
   \left(\mathbf{\Pi}\cdot\nabla\right)\cdot\mathbf{v}
\end{equation}
\begin{equation}
\nabla\cdot\mathbf{B}=0
\end{equation}
\begin{equation}
\frac{\partial{\mathbf{B}}}{\partial{t}}=\nabla\times\left(\mathbf{v}
   \times\mathbf{B}\right) + \nabla\times\left(\eta\nabla\times
   \mathbf{B}\right)
\end{equation}
\begin{equation}
\frac{\rho_1}{\rho_0}=\frac{p_1}{p_0}-\frac{T_1}{T_0}
\end{equation}
\begin{equation}
\frac{s_1}{c_p}=\frac{T_1}{T_0}-\frac{\gamma-1}{\gamma}\frac{p_1}{p_0}\, .
\end{equation} 
Here, $\rho_1$, $p_1$, $T_1$, $s_1$, $\mathbf{v}$, and $\mathbf{B}$ 
refer to the density, gas pressure, temperature, entropy, velocity, 
and magnetic field perturbations, while $\rho_0$, $p_0$, $T_0$, and
$s_0$ denote the corresponding values of the zeroth order reference 
state, described in detail below. \boldmath ${\rm g}
\mbox{\unboldmath$\,= -$g}\,\hat{z}$ \unboldmath is the acceleration 
due to gravity, and is assumed to be uniform in our calculations. The
quantity $c_p$ represents the specific heat at constant 
pressure.  The viscous stress tensor $\mathbf{\Pi}$ is given by
\begin{equation}
\Pi_{ij}\equiv\mu\left(\frac{\partial v_i}{\partial x_j}+
   \frac{\partial v_j}{\partial x_i}-\frac{2}{3}\left(\nabla\cdot
   \mathbf{v}\right)\delta_{ij}\right) \;,
\end{equation}
and $\mu$, $\eta$, $K$, represent the coefficients of viscosity, 
magnetic diffusion, and thermal diffusion respectively.  

A detailed description of the numerical methodology we use in the 
solution of the anelastic MHD equations can be found in
Appendix A of \citet{fzl99}.  Briefly, the non-dimensional form of the 
equations are solved in a rectangular domain assuming periodic boundary 
conditions in the horizontal directions, and non-penetrating, stress-free 
conditions at the upper and lower boundaries.  The magnetic field 
$\mathbf{B}$ and the momentum density $\rho_0 \mathbf{v}$ are both
divergence-free, and thus each can be expressed in terms of
two scalar potentials:
\begin{equation}
  \mathbf{B}=\nabla\times\nabla\times\mathcal{B}
    \mbox{\boldmath$\hat{z}$\unboldmath}+\nabla\times
    \mathcal{J}\mbox{\boldmath$\hat{z}$\unboldmath} \, 
\end{equation} 
and
\begin{equation}
  \rho_0\mathbf{v}=\nabla\times\nabla\times\mathcal{W}
    \mbox{\boldmath$\hat{z}$\unboldmath}+\nabla\times
    \mathcal{Z}\mbox{\boldmath$\hat{z}$\unboldmath} \, . 
\end{equation}
These potentials, along with the other dependent variables of the problem, 
are spectrally decomposed in the horizontal Cartesian directions.  
The Fourier variables are discretized with respect to the vertical 
direction, and the vertical derivatives are approximated by fourth-order, 
centered differences.  A semi-implicit method is then used to time-advance 
the five discretized scalar equations for the Fourier variables.  Using 
operator splitting, the second-order Adams-Bashforth scheme is applied to the 
advection terms, and the second-order Crank-Nicholson scheme is applied to the 
diffusion terms (see \citealt{p86} for a general discussion
of these methods).

To investigate the dynamics of flux tube fragmentation, and how this 
process depends on the initial state and eventual geometry of the tube, 
we carried out a total of $16$ simulations.  In each case, an ideal gas
of $\gamma=5/3$ is assumed; the reference state is taken to be an
adiabatically stratified polytrope of index $m=1.5$ (related to
$\gamma$ by $m=1/(\gamma-1)$); and $\mu$, $\eta$, 
and $\rho_0 K$ are assumed constant throughout the simulation domain.  
The diffusive parameters enter into the calculation via the Reynolds 
number ($R_\mathrm{e}\equiv[\rho][z][v]/\mu$), the magnetic Reynolds 
number ($R_\mathrm{m}\equiv[z][v]/\eta$), and the Prandtl number 
($P_\mathrm{r}\equiv\mu/(K\rho_0)$).  The density and temperature scales
($[\rho]$ and $[T]$) are defined as the density and temperature of 
the reference state at the bottom of the simulation domain ($\rho_r$ 
and $T_r$), and the length scale $[z]$ is defined as the pressure 
scale height of the reference state at that same location ($H_r=r_\ast 
T_r/g$, where $r_\ast\equiv R/\overline{\mu}$, and $R$ and 
$\overline{\mu}$ are the ideal gas constant and mean molecular 
weight respectively).  The velocity scaling $[v]$ is given as the 
characteristic Alfv\'en speed along the axis of the initial magnetic 
flux tube.  For each simulation, both $R_\mathrm{e}$ and $R_\mathrm{m}$ 
are set to $3500$, and $P_\mathrm{r}$ is set to unity.  

Each run begins with a static, cylindrical magnetic flux tube embedded 
in a polytropic, field-free, reference state.  The vertical domain of 
each simulation spans 5.147 pressure scale heights (or 3.088 density 
scale heights).  The tube initially has the form:
\boldmath
\begin{equation}
\mathrm{B}=\mbox{\unboldmath$B_\theta(r)$}\,\hat{\theta}+ 
   \mbox{\unboldmath$B_x(r)$}\,\hat{x} \;,
\end{equation}
\unboldmath
where
\begin{equation}
B_x(r)=B_0 e^{-r^2/a^2} 
\end{equation}
and
\begin{equation}\label{eqtwist}
B_\theta(r)=\frac{q}{a}rB_x(r) \;.
\end{equation}
Here, $B_\theta(r)$ denotes the azimuthal component of the field in 
the tube's cross-section, and $B_x(r)$ refers to the axial component 
(which lies perpendicular to \boldmath ${\rm g}\mbox{\unboldmath$\,= 
-$g}\,\hat{z}$ \unboldmath along the Cartesian direction \boldmath 
$\hat{x}$ \unboldmath ).  Both are given as functions of $r$, 
the radial distance to the central axis in the tube's cross-section.  
For each simulation, the initial size of the flux tube, $a$ (defined 
as the FWHM of $B_x(r)$), is set to $0.1H_r$, and the magnetic field 
perturbations are scaled to the initial strength of the axial field 
at tube center ($B_0$).  The distance over which a field line
rotates once around the axis of the tube is given by $2\pi a/q$, where 
$q$ is the non-dimensional twist parameter of equation~(\ref{eqtwist}).
Note that this definition of $q$ differs from that of \citet{llf96}
by a factor of tube width, $a$.  Both $q$ and $a$ are assumed constant, 
so that the initial rate of field line rotation per unit length along the 
tube ($q/a$ for length scale $a$) remains fixed.  To investigate how the 
tube's initial twist impacts the amount of fragmentation apparent during 
its rise, we consider three representative values of the twist parameter: 
$q=1/4$, $q=3/16$, and $q=1/8$.

Our goal is to model the dynamics of an emerging $\Omega$-loop.  However,
due to finite computational resources, we cannot evolve each flux tube
self-consistently from an initial state of force balance (eg. 
\citealt{cms95,ff96,f99}).  We therefore introduce an ad-hoc, 
entropy perturbation at $t=0$ that causes the tube to rise and emerge
in the shape of an $\Omega$-loop.  We argue that the physics of the 
hydrodynamic interaction of the rising loop with its environment
depends primarily on the geometry of the loop and its velocity field
rather than how it arrived at its $\Omega$-loop configuration.  The 
initial entropy perturbation is of the form 
\begin{equation}\label{eqentr}
s_1=S_0e^{-r^2/a^2}\left(e^{-\left(x-L/2\right)^2/
   \mathcal{L}L}-\frac{1}{2}\right) \, .
\end{equation}
Here, $L$ refers to the initial length of the tube (which 
corresponds to the extent of the domain in the \boldmath $\hat{x}$
\unboldmath direction), and $S_0$ denotes the relative amplitude of the
perturbation (taken to be unity in the dimensionless units of the
code, where the unit of entropy is $[s]=c_p[v]^2/(r_\ast[T])$). With 
appropriate choices of $L$ and length scale $\mathcal{L}$, this initial 
condition has the effect of ``pinning-down'' the ends of the flux tube, 
while allowing the central portion to rise.  For example, the run labeled
SL1 in Table~\ref{tbl-1} has an initial acceleration due to buoyancy of 
$0.97$ (in naturalized units of $B_0^2/(8\pi H_r\rho_r)$) at the center 
of the tube, and a buoyancy contribution of $-0.03$ at each end.  By varying
the parameters in equation~(\ref{eqentr}), we can investigate how the 
fragmentation process is affected by the geometry of a rising $\Omega$-loop.

Table~\ref{tbl-1} lists the $q$, $L$ and $\mathcal{L}$ parameter space that
is explored in each of the $16$ simulations, and assigns labels to each 
run. For convenience, Table~\ref{tbl-1} also shows the total number of field
line rotations along the finite length of the tube, $Q$. Each horizontal 
flux tube is initially positioned near the bottom of the computational domain 
at $z_0=0.1875 \, z_{\mathrm{max}}$ ($z_{\mathrm{max}}$ denotes the maximum 
vertical height of the domain).  For the three-dimensional runs, the labeling 
convention consists of two letters (``L'' or ``S'') followed by a number 
($0$, $1$, $2$, or $3$).  The numbers denote the degree of magnetic field 
line rotation about the central axis of the tube (lower numbers imply a 
lesser amount of twist).  The first letter of the label refers to the extent 
of the computational domain in the \boldmath $\hat{x}$ \unboldmath direction 
(the ``L'' in this case refers to a ``l''arge 512 zone domain, and 
the ``S'' stands for a ``s''mall 256 zone domain).  The second letter 
of the label describes the eventual radius of curvature at the apex of 
the rising tube.  If ``L'' is used, the run upon completion has a 
relatively ``l''arge radius of curvature at the apex, and the 
$\Omega$-loop spans most of the computational domain in the 
\boldmath $\hat{x}$ \unboldmath direction. Otherwise, ``S'' is used to 
denote a ``s''horter $\Omega$-loop; one which exhibits a smaller radius of 
curvature at its peak, and spans a smaller portion of the box length
(this is accomplished by varying the parameter $\mathcal{L}$ in
equation~[\ref{eqentr}] in a fixed computational box).  
Labels with only one letter refer to the two-dimensional limiting cases.  
Note that it is the number of zones in the \boldmath $\hat{x}$ \unboldmath 
direction that is changed between different cases, and \emph{not} the grid 
resolution.

\section{Results}\label{results} 

In general, the magnetic field distribution in three dimensions
can become quite complex.  We find it useful to describe the field
more intuitively in terms of the evolution and possible fragmentation of 
our initial magnetic flux tube.  To accomplish this, we specify a
new coordinate system based on the path of the flux tube through our 
Cartesian domain.  We define the path of the tube in terms of 
the Cartesian variable $x$ at any given time during a simulation.
Setting up this new coordinate system and its basis vectors is a 
multi-step process.  We first consider the magnetic field weighted 
moments of the position within vertical slices through relevant 
regions of the computational domain:
\begin{eqnarray}
   \overline{z}(x)\!\!&\equiv &\!\!\frac{1}{\Phi (x)}\int
      \!\!\!\int z\,|\mathbf{B}(x,y,z)|\,dy\,dz \nonumber \\ 
   \overline{y}(x)\!\!&\equiv &\!\!\frac{1}{\Phi (x)}\int
      \!\!\!\int y\,|\mathbf{B}(x,y,z)|\,dy\,dz \, ,
\end{eqnarray}
where
\begin{equation}
  \Phi (x)\equiv\int\!\!\!\int |\mathbf{B}(x,y,z)|
     \,dy\,dz \;.
\end{equation}
Then it is natural to define a path given by the vector
\boldmath
\begin{equation}
  {\rm r}\mbox{\unboldmath $_0(x)\,\equiv\, x$ \boldmath}\!\hat{x}
     +\mbox{\unboldmath $\overline{y}(x)$ \boldmath}\!\hat{y}
     +\mbox{\unboldmath $\overline{z}(x)$ \boldmath}\!\hat{z}
\end{equation}
\unboldmath
and to construct the Frenet tangent vector along this path, 
\begin{equation}
   \mbox{\boldmath $\hat{\ell}$\unboldmath}_0\,\equiv\,\left( 1+\left(
      \frac{d\overline{y}}{dx}\right)^2+\left(\frac{d\overline{z}}
      {dx}\right)^2\right)^{-1/2}\left(\mbox{\boldmath $\hat{x}$
      \unboldmath}\!+\frac{d\overline{y}}{dx}\mbox{\boldmath $\hat{y}$
      \unboldmath}\!+\frac{d\overline{z}}{dx}\mbox{\boldmath $\hat{z}$
      \unboldmath}\right)\;.
\end{equation}

The path traced by $\mathbf{r}_0$ is only a first approximation
to the path of the flux tube.  To establish the actual path of 
the tube, we calculate the magnetic field weighted
position along a plane normal to \boldmath $\hat{\ell}$\unboldmath$_0$ 
passing through a given point along $\mathbf{r}_0$. This plane is defined 
by the equation \boldmath $\hat{\ell}$\unboldmath$_0\cdot(\mathbf{r}-
\mathbf{r}_0)=0$, and its surface can be parameterized by 
$\mathbf{r}=u(v,w)\,$\boldmath$\hat{x}$\unboldmath$+v\,
$\boldmath$\hat{y}$ \unboldmath$+w\,$\boldmath$\hat{z}$\unboldmath ; 
where $u(v,w) \equiv (\ell_y(\overline{y}-v)+\ell_z(\overline{z}-w))
/\ell_x+x$, and the $\ell_i$'s refer to the Cartesian components 
of \boldmath $\hat{\ell}$\unboldmath$_0$.  With the area element along 
this surface given by $dS=|\partial \mathbf{r}/\partial v\times\partial
\mathbf{r}/\partial w|\,dv\,dw=(\hat{\mathbf{x}}\,\cdot\,$\boldmath$
\hat{\ell}$\unboldmath$_0)^{-1}\,dv\,dw$, the total unsigned magnetic
flux across the plane can be expressed as $\Phi_S \equiv
\int_S |\mathbf{B}(v,w)|\,dS$.  We now define a new set of moments:
\begin{eqnarray}\label{moments}
  \overline{v}\!\!&\equiv &\!\!\frac{1}{\Phi_S}\int\!\!\!\int
    v\,|\mathbf{B}(v,w)|\,\mbox{$($\boldmath $\hat{x}\cdot
    \hat{\ell}$\unboldmath$_0)^{-1}$}\,dv\,dw \nonumber \\
  \overline{w}\!\!&\equiv &\!\!\frac{1}{\Phi_S}\int\!\!\!\int
    w\,|\mathbf{B}(v,w)|\,\mbox{$($\boldmath $\hat{x}\cdot
    \hat{\ell}$\unboldmath$_0)^{-1}$}\,dv\,dw \, .
\end{eqnarray}
Along with $\overline{u}=u(\overline{v},\overline{w})$, 
these points are used to define the path of a magnetic flux 
tube in a given region,
\boldmath
\begin{equation}\label{path}
   r\mbox{ \unboldmath$\,\equiv\,\overline{u}\,$\boldmath}\hat{x}
      \mbox{ \unboldmath $+\,\overline{v}\, $\boldmath}\hat{y}
      \mbox{ \unboldmath $+\,\overline{w}\, $\boldmath}\hat{z}
      \mbox{\unboldmath$\, . $\boldmath}
\end{equation}
\unboldmath

The frame of reference of the tube is then given by the Frenet 
tangent, normal, and binormal vectors along \boldmath $r$\unboldmath : 
\boldmath$\hat{\ell}$\unboldmath$\,\,=d$\boldmath$r$\unboldmath$/ds\,$ 
(where $ds$ is the infinitesimal path length along \boldmath 
$r$\unboldmath ), \boldmath $\hat{n}$\unboldmath$\,=
\kappa^{-1}d$\boldmath$\hat{\ell}$\unboldmath$/ds\,$, and 
\boldmath$\hat{b}$\unboldmath$\,\,=\,\,$\boldmath$\hat{\ell}$\unboldmath$
\,\times\,$\boldmath$\hat{n}\,$\unboldmath, respectively.  
The tube's curvature at a given point along \boldmath $r$ \unboldmath 
is simply $\kappa=|d$\boldmath$\hat{\ell}$\unboldmath$/ds|$. Of course,
where the curvature of the flux tube is zero, the normal (and hence
binormal) vectors are ill-defined.  However, in the course of our
analysis, we find that zero curvature occurs only at very localized 
$\kappa$ inflection points, or along horizontally oriented tubes 
(where vertical slices can be used to define a cross-sectional plane),
and thus this limitation proves inconsequential.  We note that  
our formalism is not terribly general, as it precludes
the consideration of unusual configurations such as vertically oriented
tubes, or tubes that are stacked on top of one another; however,
none of these situations are encountered in our study.

Depending on the initial degree of twist, portions of the rising 
tube will shed vortex pairs, and the magnetic flux will be 
redistributed in the tube cross-section such that much of the 
flux is located away from the tube's central axis. In some cases,
the distribution of magnetic flux no longer resembles
a single, cohesive tube; rather it has spread out and split apart
into a configuration that can best be described as two separate
flux tubes. At this point, we consider the tube to be ``fragmented''. 
A quantitative measure of this fragmentation can be obtained by first 
calculating the second moments of the magnetic field strength along 
the normal and binormal directions of the tube:
\begin{eqnarray}\label{secmom}
   <\!(n-\overline{n})^2\!>\!\!&\equiv &\!\!\frac{1}{\Phi_{S^\prime}}
      \int_{S^\prime}\left(n-\overline{n}\right)^2|\mathbf{B}(n,b)|\,
      d{S^\prime} \nonumber \\
   <\!(b-\overline{b})^2\!>\!\!&\equiv &\!\!\frac{1}{\Phi_{S^\prime}}
      \int_{S^\prime}\left(b-\overline{b}\right)^2|\mathbf{B}(n,b)|\,
      d{S^\prime} \, .
\end{eqnarray}
Here, $\mathbf{r}=n$\boldmath $\hat{n}$ \unboldmath $+\,b$\boldmath
$\hat{b}\,$ \unboldmath spans the two-dimensional space of the plane 
normal to \boldmath $\hat{\ell}$ \unboldmath, $d{S^\prime}$ is 
the area element in that plane, and $\Phi_{S^\prime}\equiv\int_{S^\prime}
|\mathbf{B}(n,b)|\,d{S^\prime}$ is the corresponding unsigned 
flux (the \boldmath $\ell$ \unboldmath dependence of the 
variables is implicitly understood). In the above equations, 
$\overline{n}\equiv(1/\Phi_{S^\prime})\int_{S^\prime} n\,|\mathbf{B}(n,b)
|\,d{S^\prime}$ and $\,\overline{b} \equiv (1/\Phi_{S^\prime})
\int_{S^\prime} b\,|\mathbf{B}(n,b)|\,d{S^\prime}$ represent the 
first moments of the field distribution in this geometry.  Due to 
the symmetry of our particular problem, the relative deformation of
the tube along its cross-section can be represented by the 
ratio of the binormal to normal second moments:
\begin{equation}
   f=\frac{<\!(b-\overline{b})^2\!>}
      {<\!(n-\overline{n})^2\!>} \, .
\end{equation}
The total spread of the field distribution, $\sigma^2=(1/\Phi_{S^\prime})
\int_{S^\prime} \left((n-\overline{n})^2+(b-\overline{b})^2\right)|
\mathbf{B}(n,b)|\,d{S^\prime}$, can be used as a measure of how far the
tube has dispersed in the cross-sectional plane. 
  
\begin{figure}
\epsfig{file=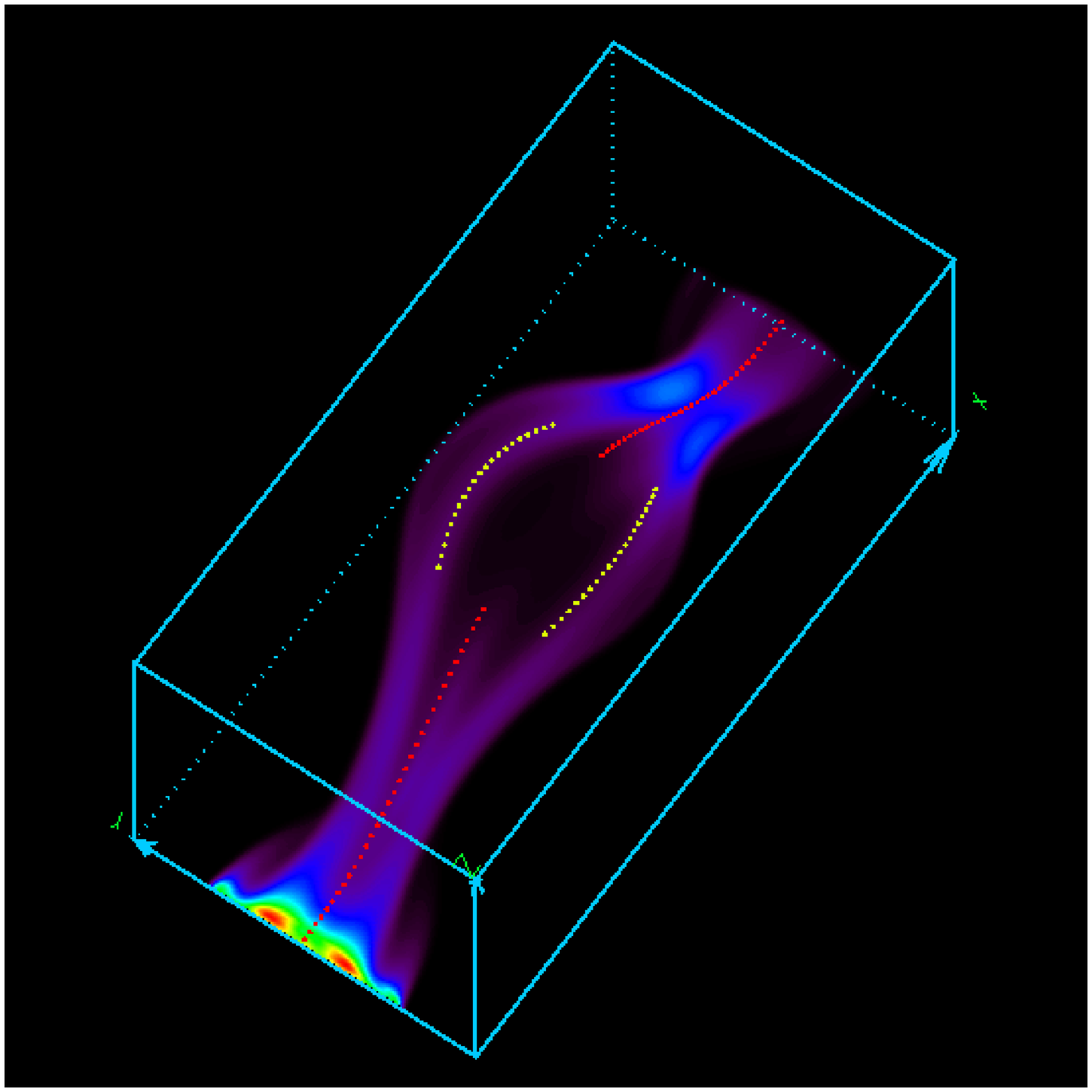,scale=.70,bbllx=-40,bbury=760}
\caption[f1.eps]{A volume rendering of $B^2$ for the last timestep
of run SL0. The red dotted line denotes the central axis of the portion
of the $\Omega$-loop which has not fragmented. The yellow dotted lines
denote the paths of each distinct fragment in the region of the loop
that is considered to be fragmented.  The entire rectangular computational
domain is shown.\label{fig1}}
\end{figure}

\begin{figure}
\epsfig{file=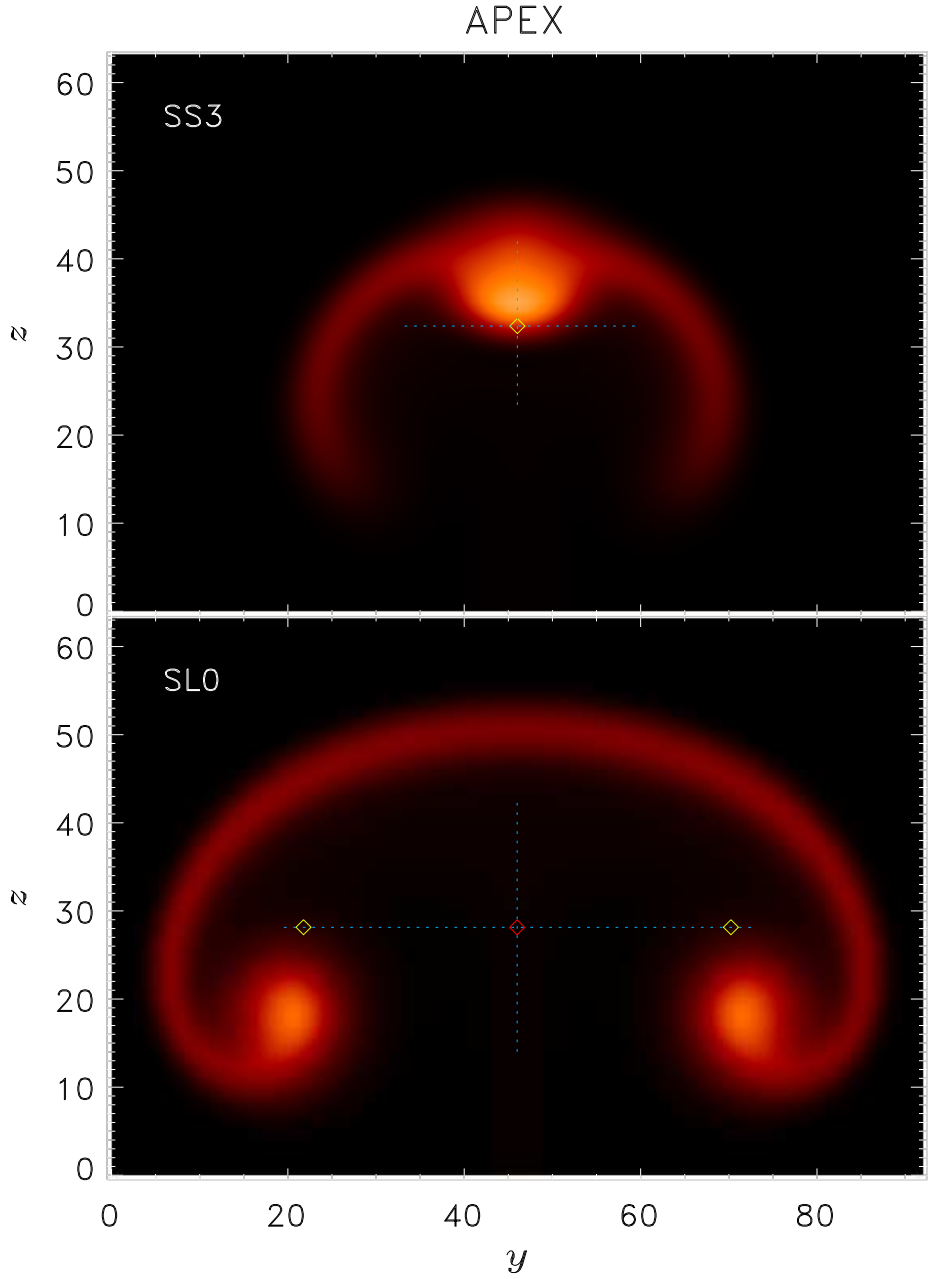,scale=.9, bbllx=-20,bbury=550}
\caption[f2.eps]{The apex cross-sections of the $\Omega$-loops shown
in Figure~\ref{fig1} (bottom panel) and Figure~\ref{fig3} (top panel).
The symbols denote the positions of the tube axes, and the lengths of
the blue dotted lines represent the second moments of the field
distribution centered on the tube centroid (see
equation~[\ref{secmom}]).  These lines serve as a visual measure 
of the degree of fragmentation.  The $x$ and $y$ scale is given in 
terms of computational zones.  \label{fig2}}
\end{figure}

\begin{figure}
\epsfig{file=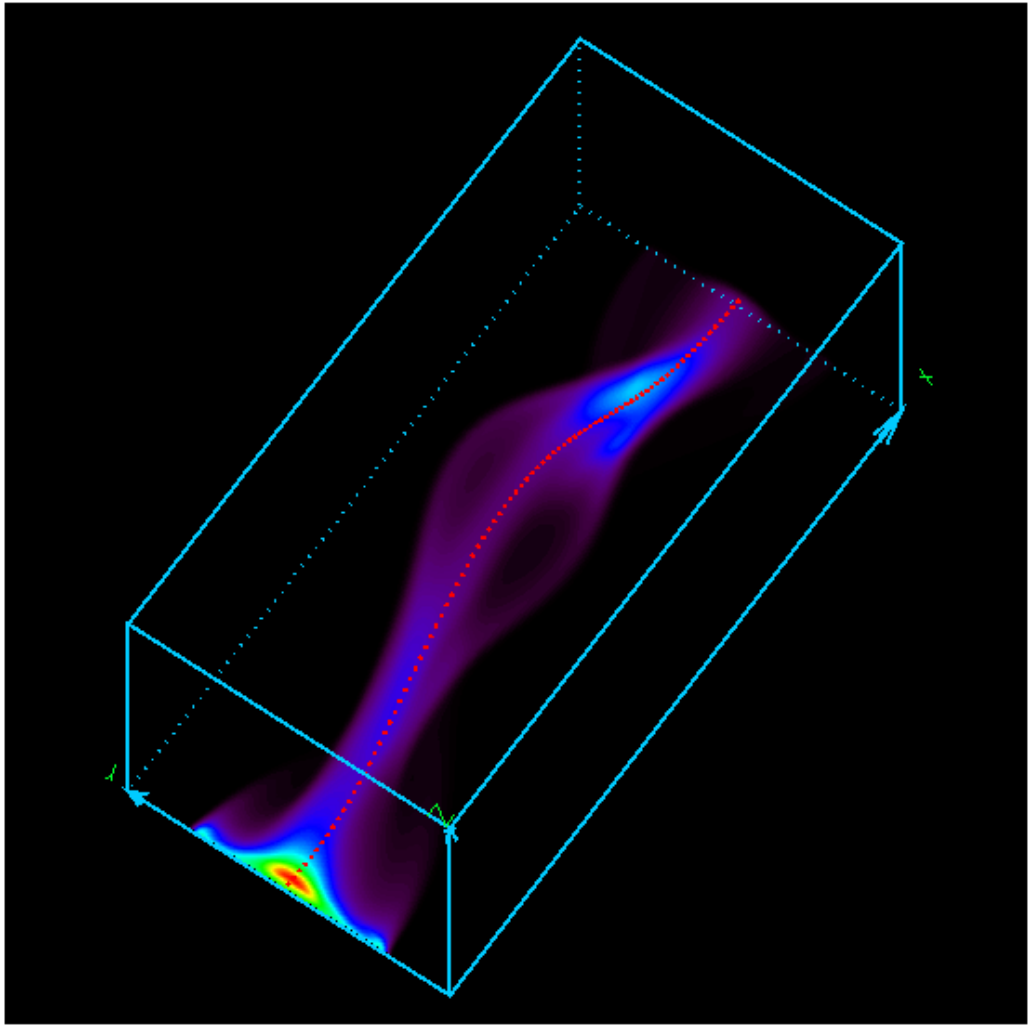,scale=.70,bbllx=-40,bbury=775}
\caption[f3.eps]{A volume rendering of $B^2$ for the last timestep
of run SS3.  The red dotted line denotes the central axis of the
non-fragmented flux tube. \label{fig3}}
\end{figure}

From an examination of numerous simulations, we empirically find that
a magnetic flux tube can be considered to have split apart when its 
deformation, $f$, exceeds 1.5.  We therefore use $f$ as our measure
of the degree of fragmentation of the tube, consistent with the 
results of \citet{s79} and \citet{lfa96}.  If $f>1.5$, then 
the path of each individual flux tube fragment is calculated using
equations~(\ref{moments}) and (\ref{path}) over subdivided portions
of the surface $S$.  Since there is a fairly high degree of 
symmetry in our runs, the surface can be divided into two parts that 
lie on either side of a line defined by the Frenet normal.  Thus, 
the regions of integration are easily determined.  Figure~\ref{fig1} is a 
volume rendering of the magnetic field strength for the final timestep of
run SL0, and shows the geometry of a tube which has fragmented as
it rises toward the surface.  The field strength distribution is visualized 
by casting parallel rays through the semi-transparent volume, and 
calculating the two-dimensional projection onto the viewing plane 
(IDL's ``voxel\_proj'' routine).  The red dotted line denotes the 
central axis of the tube through the volume.  Where the tube has
fragmented, yellow dotted lines trace the paths of each fragment.
Note that the axis of each tube fragment does not coincide exactly 
with their centers of vorticity (where the field strength is locally 
concentrated); this offset depends on the amount of flux that resides 
in a thin ``sheath'' that extends from the forefront of the rising tube 
to the outer edge of each vortex pair.  These features can be seen in a 
cross-section at the apex of this $\Omega$-loop in the bottom frame of 
Figure~\ref{fig2}.  In this case, the magnetic sheath is revealed 
as a relatively weak concentration of flux in the shape of a thin, 
semi-circular arch that lies above the primary concentrations of flux 
located in the trailing eddies.  The yellow symbols denote the location 
of the central axis of each fragment, and the lengths of the blue dotted 
lines represent the second moments of the field distribution (given by 
equation~[\ref{secmom}]) at the centroid of the tube (note that at the 
apex of the loop, \boldmath $\hat{n}$ \unboldmath and \boldmath $\hat{b}$ 
\unboldmath approximately correspond to the Cartesian 
directions \boldmath $\hat{z}$ \unboldmath and \boldmath $\hat{y}$ 
\unboldmath respectively).   The top panel of Figure~\ref{fig2} shows 
the apex cross-section of the flux tube shown in Figure~\ref{fig3} 
(the final timestep of run SS3).  This is an example of a tube which, 
by our definition, has not yet fragmented as it approaches the 
photospheric boundary.  
 
\subsection{How Fragmentation Depends on Tube Curvature and Twist}

To investigate how the degree of fragmentation at the apex 
of a rising, magnetic flux tube depends on its curvature at that 
point, we consider five separate runs: SS1, SL1, LS1, LL1, and L1.  
In each case, the initial twist parameter $q$ is taken to be $1/8$; 
that is, the simulations begin with identical, weakly-twisted 
horizontal flux tubes positioned at the same depth near the bottom of 
the computational domain.  Effectively, the difference between each 
run is the length in the axial direction of the rising portion of the 
flux tubes. The changes in this length scale lead to substantive 
differences in the apex curvature of the $\Omega$-loop 
as it rises through the stratified atmosphere.  
Run L1 (the two-dimensional limiting case) represents an infinitely 
long horizontal tube whose apex curvature remains zero throughout 
its rise.  Conversely, run SS1 represents the shortest length scale we 
have considered, and thus refers to the run with the highest apex 
curvature as the tube approaches the photosphere.

\begin{figure}
\epsfig{file=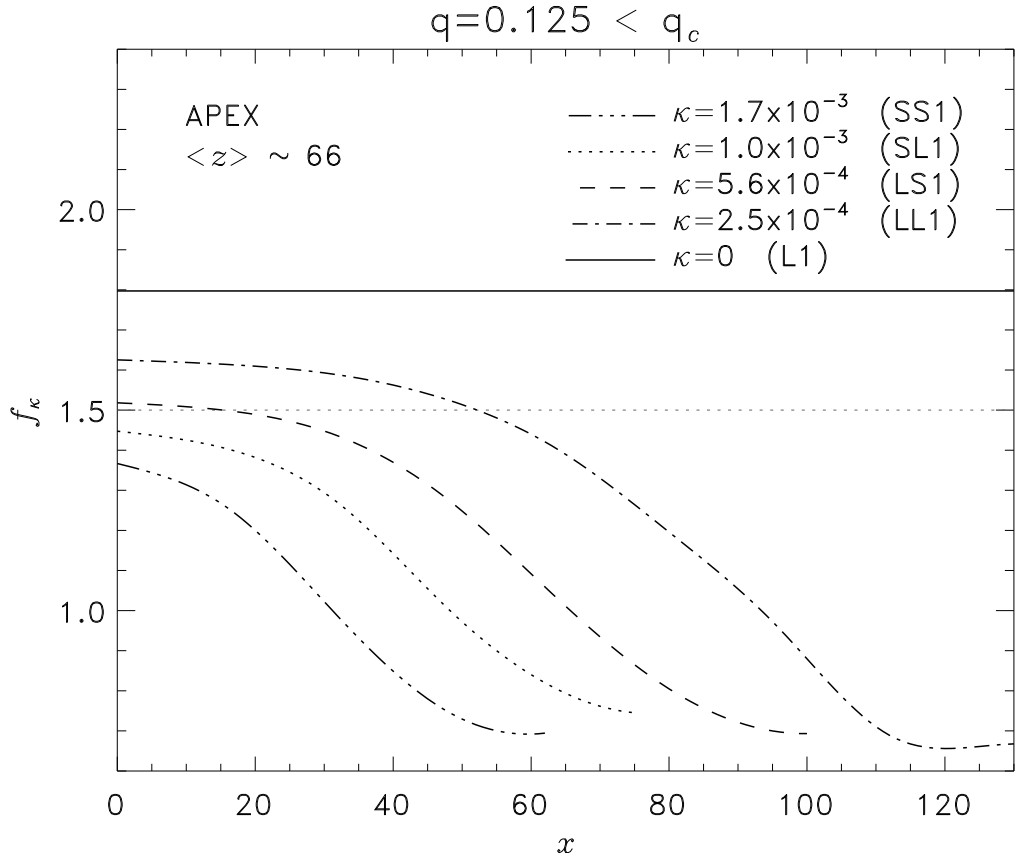,scale=1.,bbllx=35,bbury=375}
\caption[f4.eps]{Fragmentation as a function of curvature
for $q=1/8$. The $x$ scale is given by the number of computational
zones from each tube's apex, and $<\!\!z\!\!>$ is the average height of
the central axis of the tubes (given in terms of the number of computational
zones from the lower boundary). For clarity, the $f$ values near the
horizontal boundaries of each run (past the end of the $\Omega$-loop)
are not shown.  \label{fig4}}
\end{figure}

Figure~\ref{fig4} shows the degree of fragmentation along the axis 
of each tube for these five values of apex curvature at the time 
when each tube has risen approximately half-way through the vertical 
extent of the domain. In order of decreasing curvature, the 
simulations shown are runs SS1, SL1, LS1, LL1, and L1.  The line 
$f=1.5$ represents the (somewhat arbitrary) point at which we 
consider the tube to be fragmented.  It is easy to see that as the level 
of apex curvature $\kappa$ increases, the less the tube fragments for a 
fixed value of $q$.  Figure~\ref{fig4} also shows that although
the initial value of twist was slight, three of the four simulations 
have yet to show clear signs of fragmentation at the apex of the loop.  
This suggests that in three dimensions, the amount of twist necessary to 
prevent fragmentation is substantially reduced from the two-dimensional 
value --- a point that we will return to later in this section.
Note that toward the footpoints of each loop, the value of 
$f_\kappa$ falls below $1$ (the $\kappa$ subscript is included
to emphasize the dependence of the degree of fragmentation $f$
on the tube's apex curvature $\kappa$).  This reflects the net 
elongation of the tube cross-section in the \boldmath $\hat{n}$ 
\unboldmath direction as the $\Omega$-loop expands.  

\begin{figure}
\epsfig{file=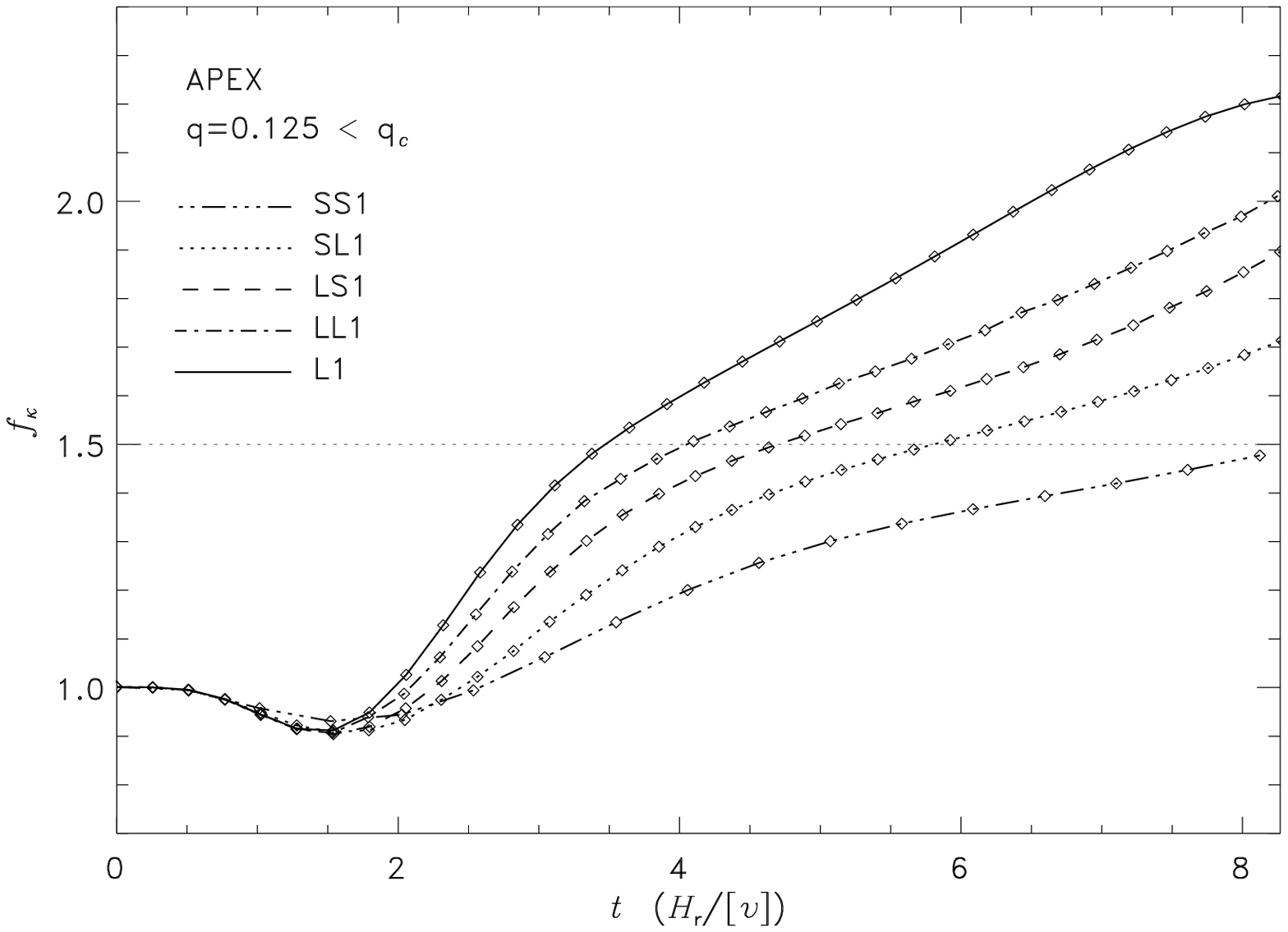,scale=1.,bbllx=35,bbury=375}
\caption[f5.eps]{Apex fragmentation as a function of time and
apex curvature for the set of runs where $q=1/8$. \label{fig5}}
\end{figure}

Figure~\ref{fig5} shows the time dependence of $f_\kappa$ at the loop 
apex for the set of runs corresponding to those of Figure~\ref{fig4}.  
A striking feature of this plot is the reduction by a factor of 
$\approx 1.5$ in the degree of fragmentation near the photospheric 
boundary between the run with zero apex curvature, and the run with 
the maximum amount of curvature.  The difference in $f_\kappa$ is large 
enough that the magnetic flux emerges as two individual fragments 
when $\kappa=0$ (run L1), yet is closer to being a single, cohesive tube 
when $\kappa\gtrsim 1.7\times 10^{-3}$ (run SS1).  Note that during the 
initial stages of the run, before the tube shows signs of fragmentation, 
$f_\kappa$ falls slightly below $1$.  This is a result of the 
vertical elongation of the apex cross-section as the tube begins to 
rise, and flux is pulled into its wake.  

\begin{figure}
\epsfig{file=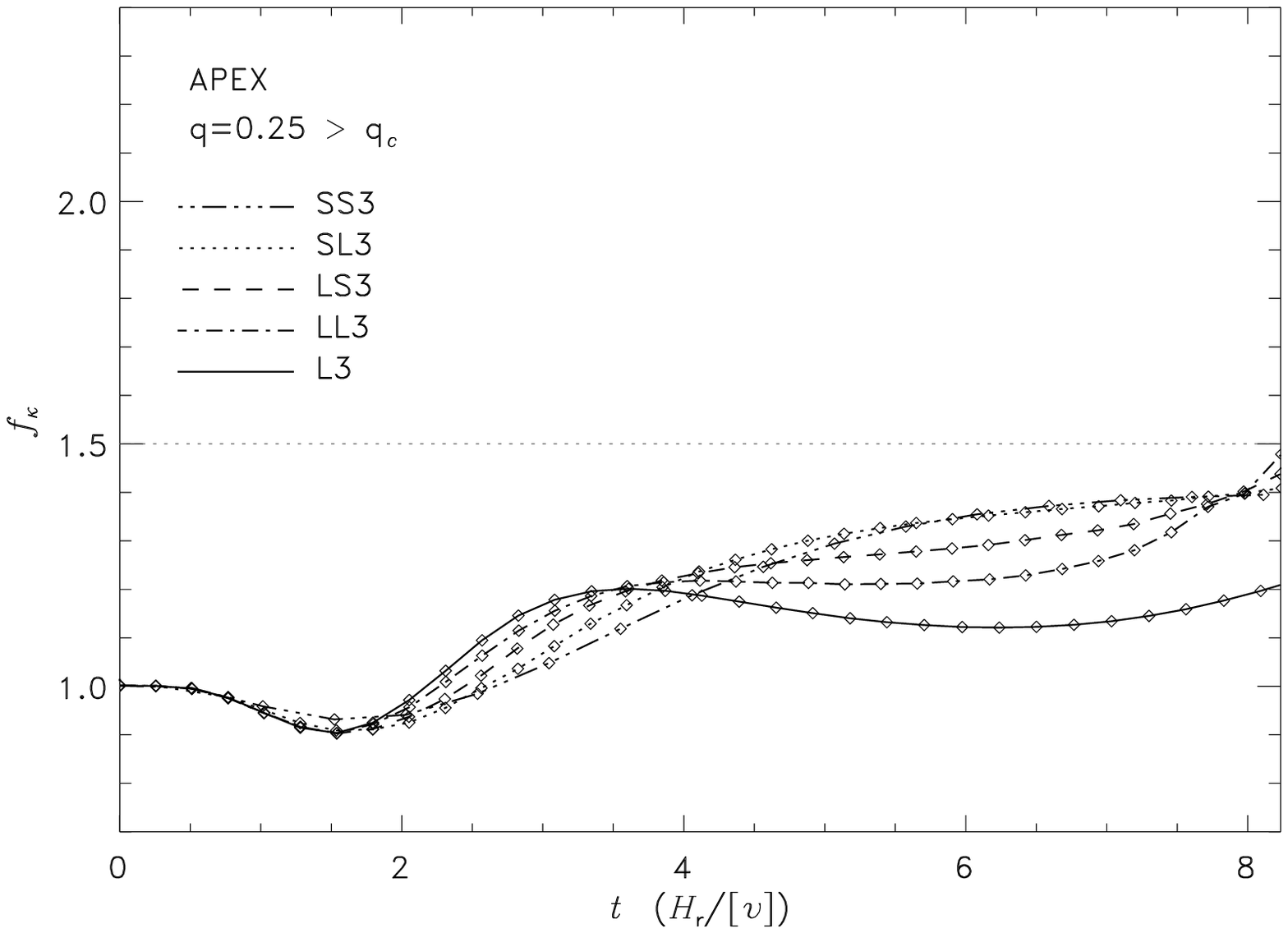,scale=1.,bbllx=35,bbury=375}
\caption[f6.eps]{Apex fragmentation as a function of time and
apex curvature for the set of runs where $q=1/4$.  \label{fig6}}
\end{figure}

\begin{figure}
\epsfig{file=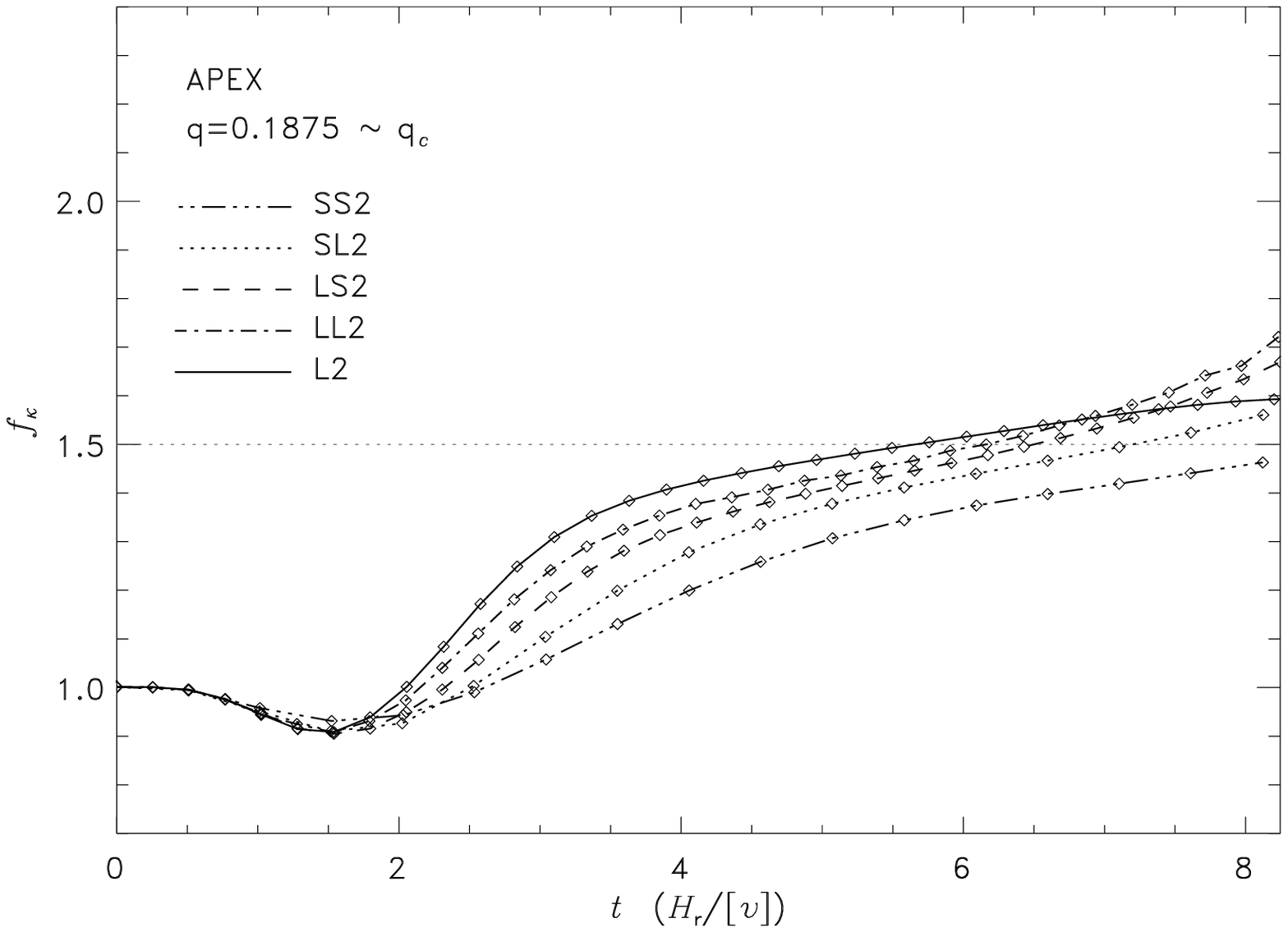,scale=1.,bbllx=35,bbury=375}
\caption[f7.eps]{Apex fragmentation as a function of time and
apex curvature for the set of runs where $q=3/16$. \label{fig7}}
\end{figure}

We find that if the initial value of the twist 
parameter $q$ is greater than a ``critical'' value, 
$q_{c}$, then the flux tube no longer emerges in a fragmented state,
($f < 1.5$) regardless of the radius of curvature of the loop (see 
Figure~\ref{fig6}).  In our simulations, $q_{c}$ is empirically 
determined to be only slightly less than $3/16$, the value chosen for 
the set of runs SS2, SL2, LS2, LL2, and L2 of Figure~\ref{fig7}. 
For the two-dimensional limiting cases, the empirically determined 
critical value is consistent with the condition that the flux tube
will split apart when the rise velocity of the tube exceeds the 
Alfv\'en speed of the azimuthal field at its edge (note that
the initial entropy perturbation in our simulations will affect
the rise speed of the tube).  However, if the initial twist of 
the flux tube is less than $q_{c}$, we find that the 
curvature of the $\Omega$-loop plays an important 
role in determining the degree of fragmentation of the tube prior to its
emergence through the photosphere --- an effect not accounted for
in previous two-dimensional studies of flux tube fragmentation 
(eg. \citealt{lfa96}, \citealt{em98}, and \citealt{fzl98}).  
A comparison of Figure~\ref{fig7} with Figure~\ref{fig5} further 
suggests that as one chooses values of $q$ progressively less than 
$q_c$, the effect of loop curvature on the degree of apex fragmentation 
becomes more pronounced.   

\begin{figure}
\epsfig{file=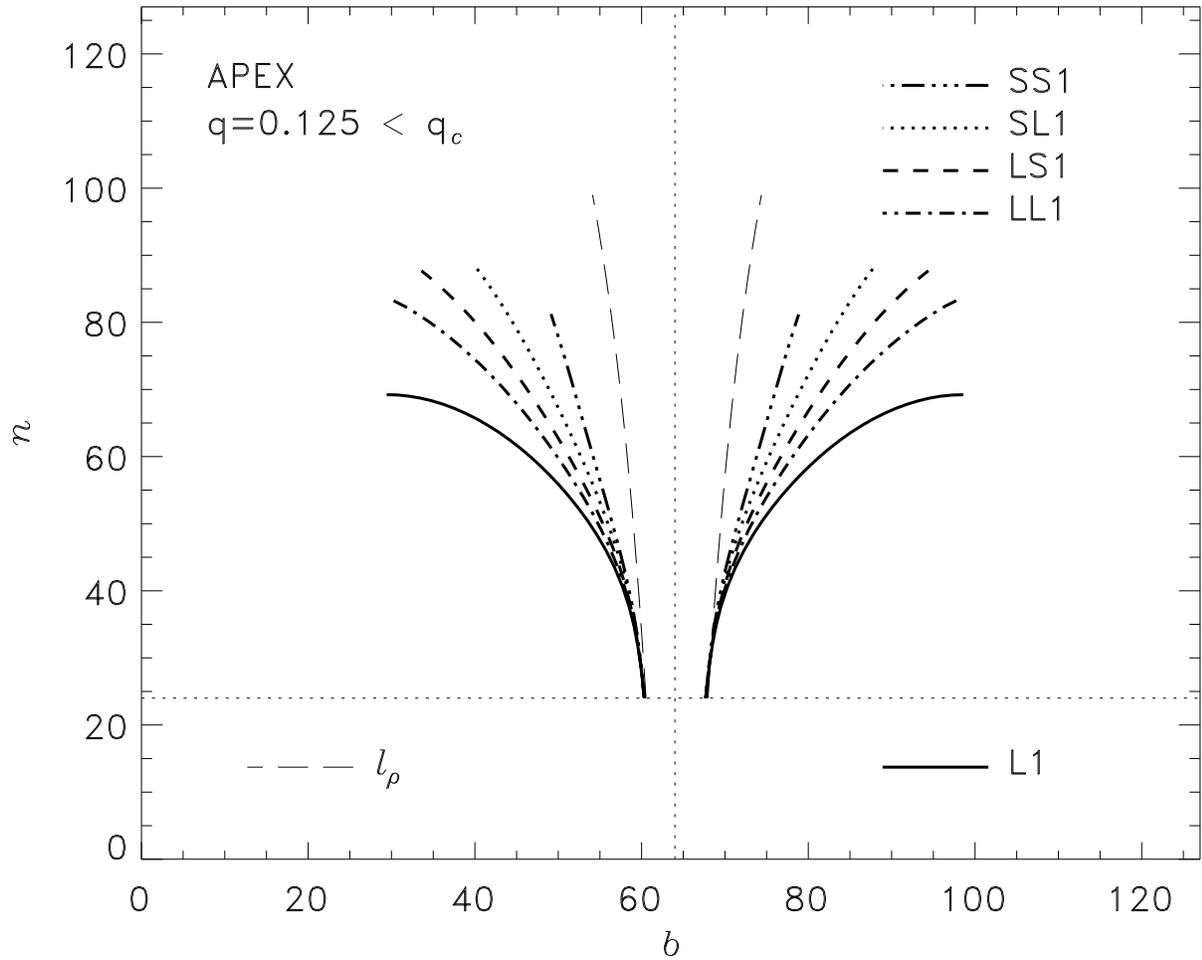,scale=1.,bbllx=65,bbury=400}
\caption[f8.eps]{Fragment trajectories at the apex of the
rising $\Omega$-loop for the set of simulations where $q=1/8$. Note
that at the loop apex, the normal $n$ and binormal $b$ directions
correspond approximately to the Cartesian directions $z$ and $y$
respectively. $\ell_\rho$ represents the trajectory expected from pure
expansion and no vortex interaction.  The axis scale is in terms of
computational zones --- an entire vertical slice of the domain is
shown.  \label{fig8}}
\end{figure}

Run L1, the two-dimensional limiting case with a relatively
small amount of initial field line twist, confirms the results
of previous two-dimensional simulations \citep{lfa96,em98,fzl98},  
which show that once fragmentation has occurred,
the two counter-rotating fragments repel one another via
forces that result from infinitely long vortex 
lines.  In two dimensions, the horizontal separation of the
fragments can be understood in terms of the hydrodynamic force
acting on an object with a net circulation \boldmath $\Gamma$
\unboldmath moving relative to a fluid with a velocity \boldmath 
$v$\unboldmath: $\mathbf{F}=-\rho_0$\boldmath$v$\unboldmath
$\times$\boldmath$\,\Gamma$\unboldmath $\,$ (see \citealt{fzl98}).  
It is the component of this ``lift'' force per unit volume acting 
along the line between the centers of vorticity of the tube fragments 
that acts to push the tubes apart.  In three dimensions, the fragments 
are finite in extent, and thus these forces are important only over a 
finite length of the loop.  Figure~\ref{fig8} shows the trajectory
of the two fragments (for the weakly-twisted case) at the tube 
apex for five different values of curvature.  As the level of 
apex curvature of the tube increases, and the effective axial 
length scale of each vortex pair decreases, we see that the total
volume integrated non-vertical component of the hydrodynamic lift 
--- and thus the fragment separation --- is reduced.

\begin{figure}
\epsfig{file=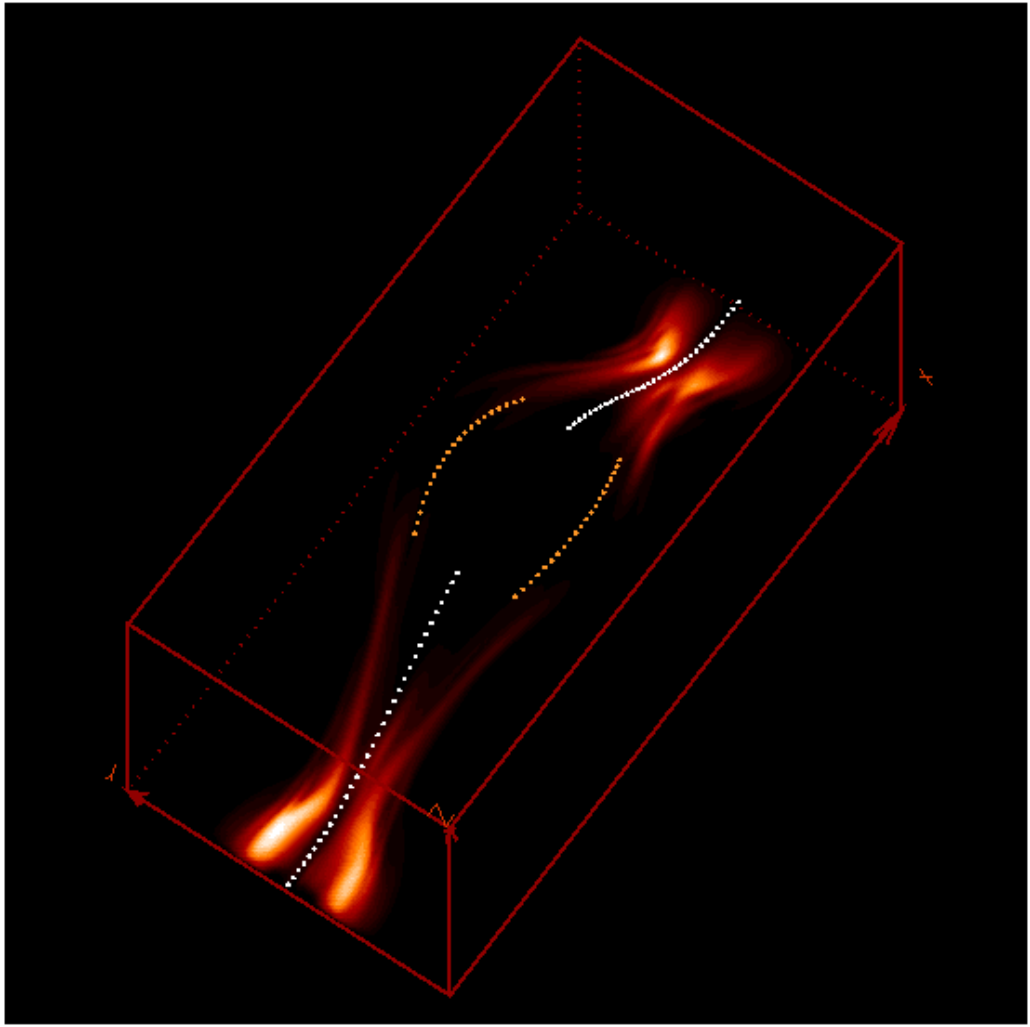,scale=.70,bbllx=-40,bbury=775}
\caption[f9.eps]{A volume rendering of the current helicity
for the last timestep of run SL0.  The dotted lines denote the
paths of both the fragmented and un-fragmented portions of the loop.
The entire rectangular computational domain is shown.
\label{fig9}}
\end{figure}

\begin{figure}
\epsfig{file=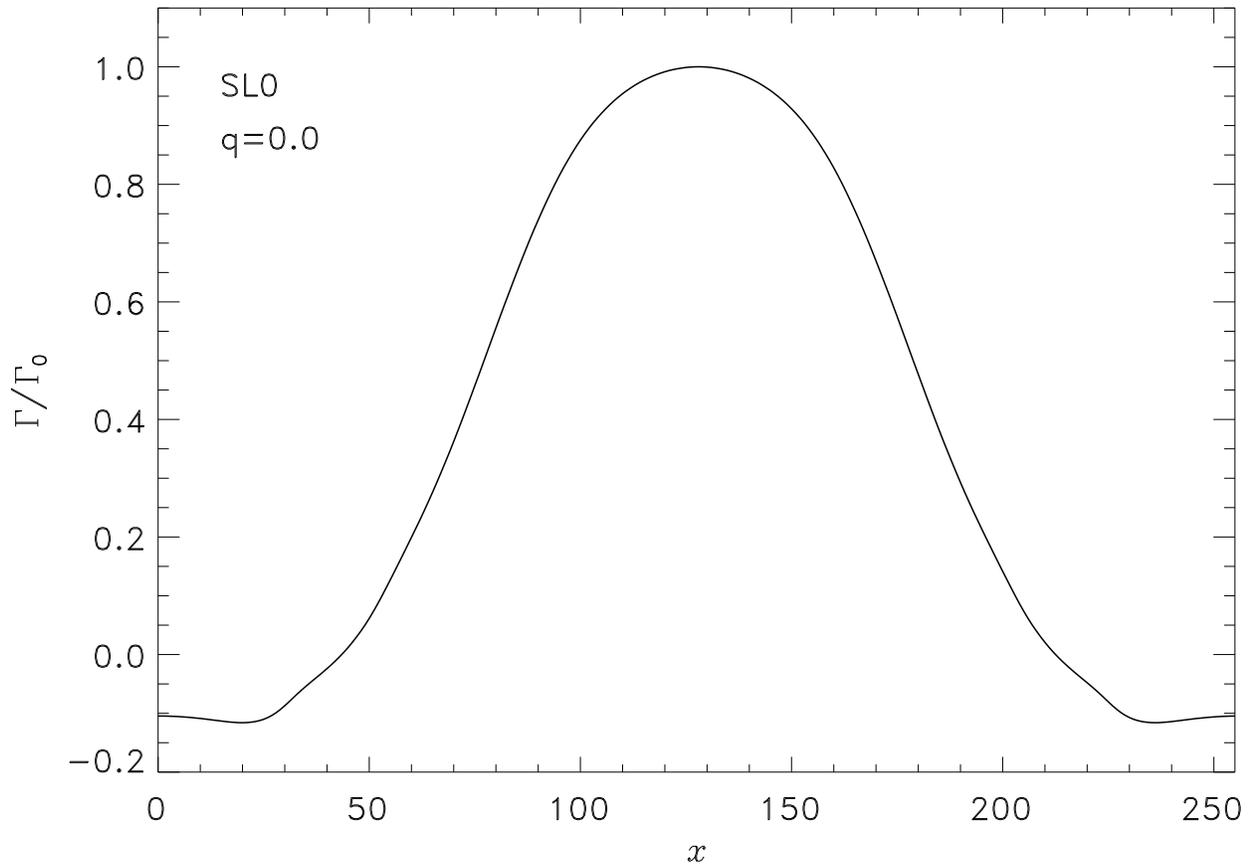,scale=1.,bbllx=25,bbury=375}
\caption[f10.eps]{The magnitude of the net circulation of an
individual tube fragment (normalized to its value at the tube's apex,
$\Gamma_0$) for the last timestep of run SL0.  \label{fig10}}
\end{figure}

\begin{figure}
\epsfig{file=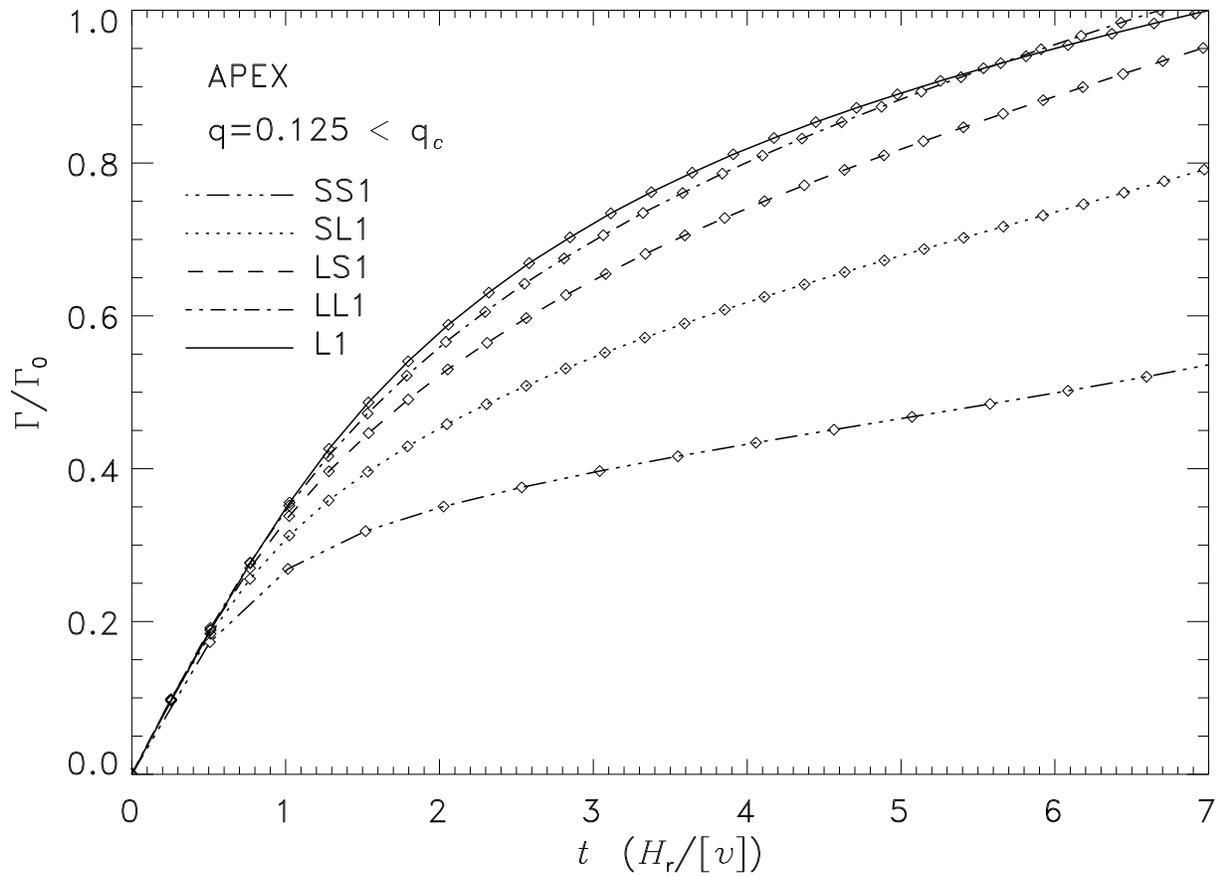,scale=1.,bbllx=30,bbury=375}
\caption[f11.eps]{The magnitude of the net circulation of
a tube fragment (normalized to the value of run L1 at $t=7$, $\Gamma_0$)
at the apex of the $\Omega$-loops as a function of apex curvature for
the runs with a relatively small amount of initial field line twist.
\label{fig11}}
\end{figure}

Field line twisting due to non-uniform rotation of the fragments
also acts to reduce the fragment separation.
To illustrate the role of field line twist in the interaction 
of the vortex pairs, we consider a case where the azimuthal 
component of the magnetic field along the tube is initially zero. 
Figure~\ref{fig9} is a volume rendering of the magnitude of the current 
helicity density ($|H_\mathrm{c}|=|\mathbf{J}\cdot\mathbf{B}|$, 
where $\mathbf{J}=\nabla\times\mathbf{B}$) for the final 
timestep of run SL0 (where $q\equiv 0$).  Since the current helicity
gives a measure of the twist of the magnetic field, Figure~\ref{fig9} 
shows that after the tube rises, spins differentially (see 
Figure~\ref{fig10}), and splits apart, the magnetic field along each 
tube fragment becomes increasingly twisted.  The magnetic forces imparted 
from the increased twist slow the rotational motion of the vortex pairs, and 
thus the tendency for the tubes to separate is further reduced.  
Figure~\ref{fig11} shows that as the apex curvature of the $\Omega$-loop 
increases, the net circulation near the apex of each fragment decreases.  
This lessens the repulsive force between fragments, and allows 
$\Omega$-loops with a high degree of apex curvature to behave more
cohesively.  This suppression of circulation in the tube fragments
was predicted by \citet{em98} and \citet{m97}, who suggested that if 
the footpoint separation of an $\Omega$-loop was small enough, the 
rotation of the vortex pairs would be suppressed.  Since it is
reasonable to assume a correlation between footpoint separation
and apex curvature, we feel that the results of our simulations
are generally consistent with this prediction.  

\subsection{Field Morphology Prior to Emergence}

There are a number of reasons why one must be cautious when comparing
our results with flux emergence observations.  First, the anelastic 
approximation becomes marginal as the magnetic flux tube approaches 
the photosphere, where densities decrease to the point that local 
sound speeds are comparable to Alfv\'en speeds in the plasma.  
Additionally, in this first generation of models we do not include 
the effects of spherical geometry or the Coriolis force. Thus, 
predicted asymmetries in the geometry and field strength asymmetry of
emerging active regions (see \citealt{msc94} and \citealt{ffd93}) 
will not be reproduced.  Further, we choose to use a simple, 
polytropic reference state rather than a more realistic convective 
background.  As a result, the effects of convective 
turbulence on the tube are neglected during its rise to 
the surface, likely over-simplifying the structure of the field.  
Nevertheless, we can make some \emph{general} predictions regarding 
the morphology of emerging magnetic flux in active regions if the field 
exists in the form of a fragmented flux tube.

\begin{figure}
\epsfig{file=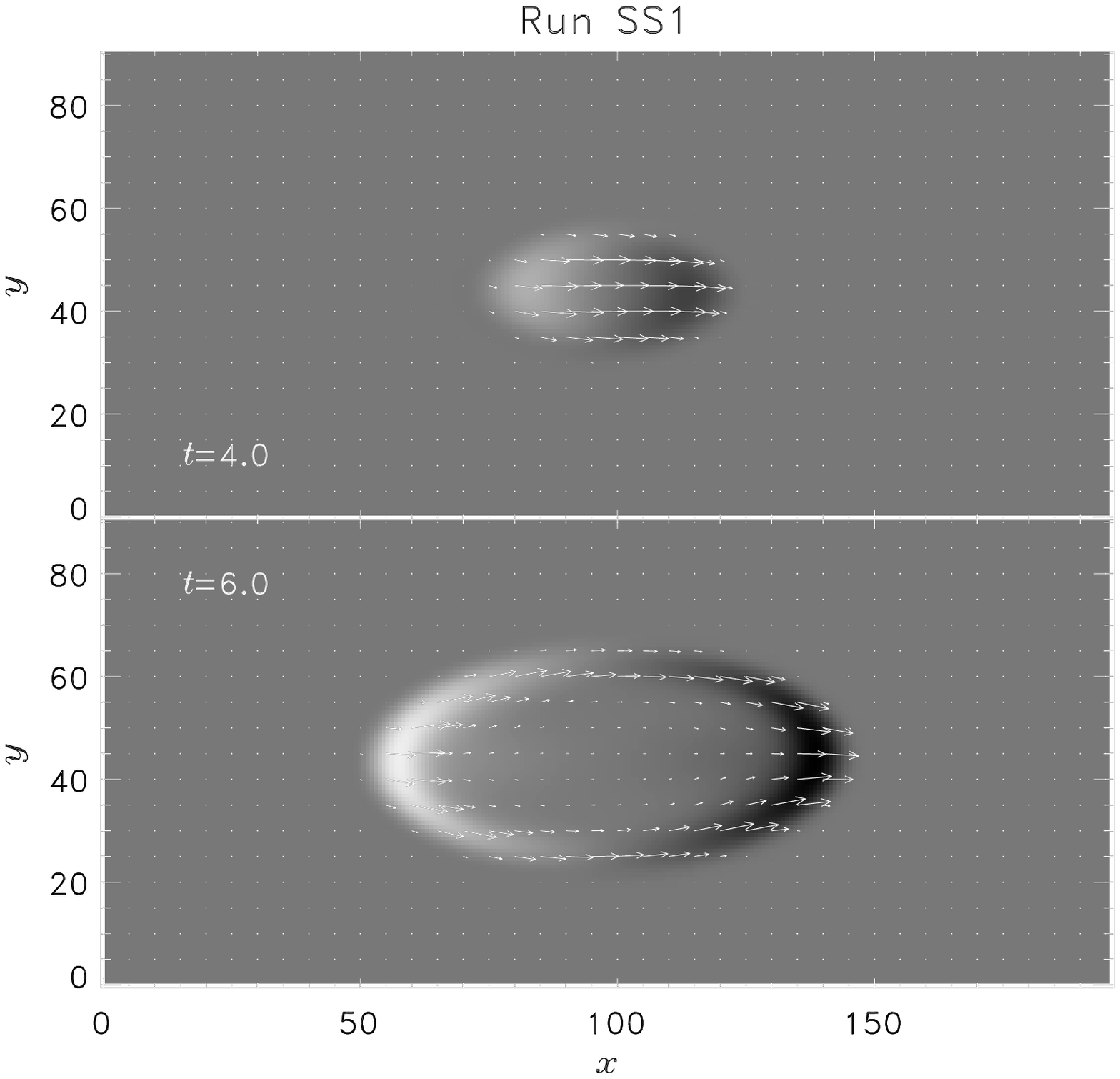,scale=.9,bbllx=75,bbury=550}
\caption[f12.ps]{Vector magnetogram images for run SS1. The
times given are in naturalized units ($t=H_r/[v]$, see text), and
the $x$ and $y$ scale are given in terms of the computational grid.
The grey-scale background represents the vertical component of the
magnetic field (positive values are indicated by the light regions), 
and the arrows represent the transverse components.
\label{fig12}}
\end{figure}

\begin{figure}
\epsfig{file=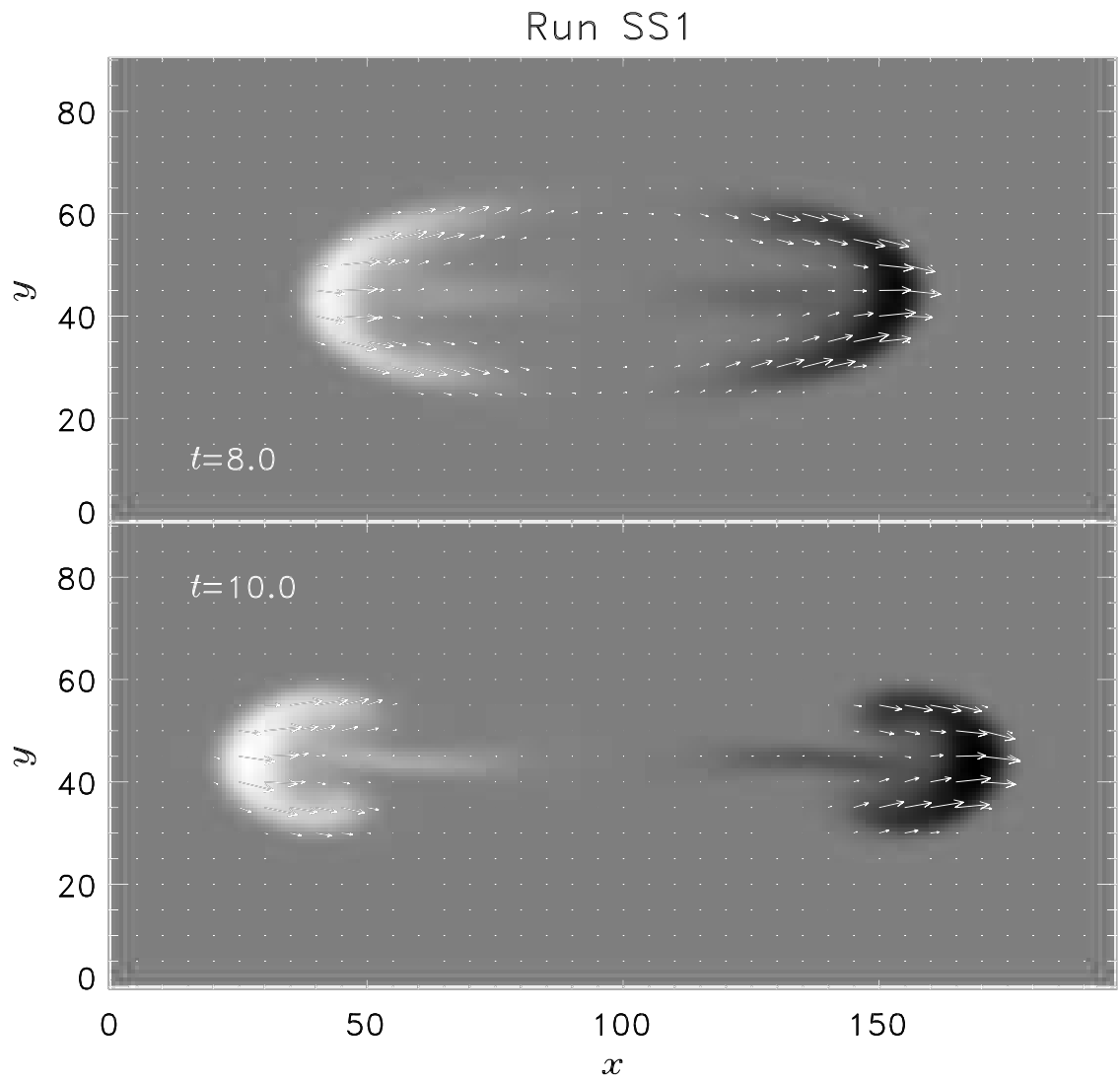,scale=.9,bbllx=75,bbury=550}
\caption[f13.ps]{Same as Figure~\ref{fig12}, for later timesteps.
\label{fig13}}
\end{figure}

\begin{figure}
\epsfig{file=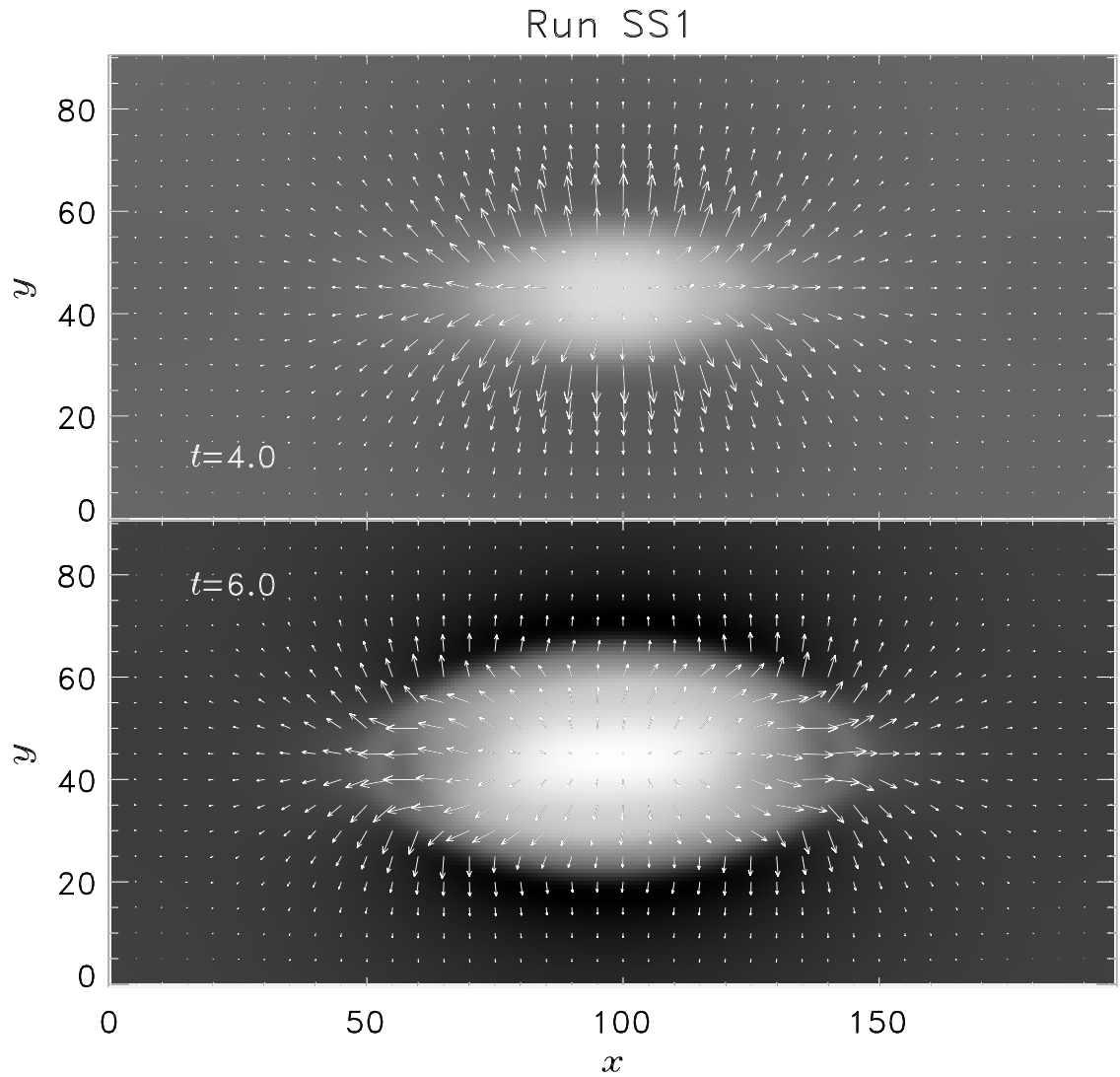,scale=.9,bbllx=75,bbury=550}
\caption[f14.ps]{The velocity field for the same horizontal slice
and times shown in Figure~\ref{fig12}.  The grey-scale background
represents the vertical component of the velocity (light regions
indicate positive values), and the arrows represent the transverse
components of the velocity field.  \label{fig14}}
\end{figure}

\begin{figure}
\epsfig{file=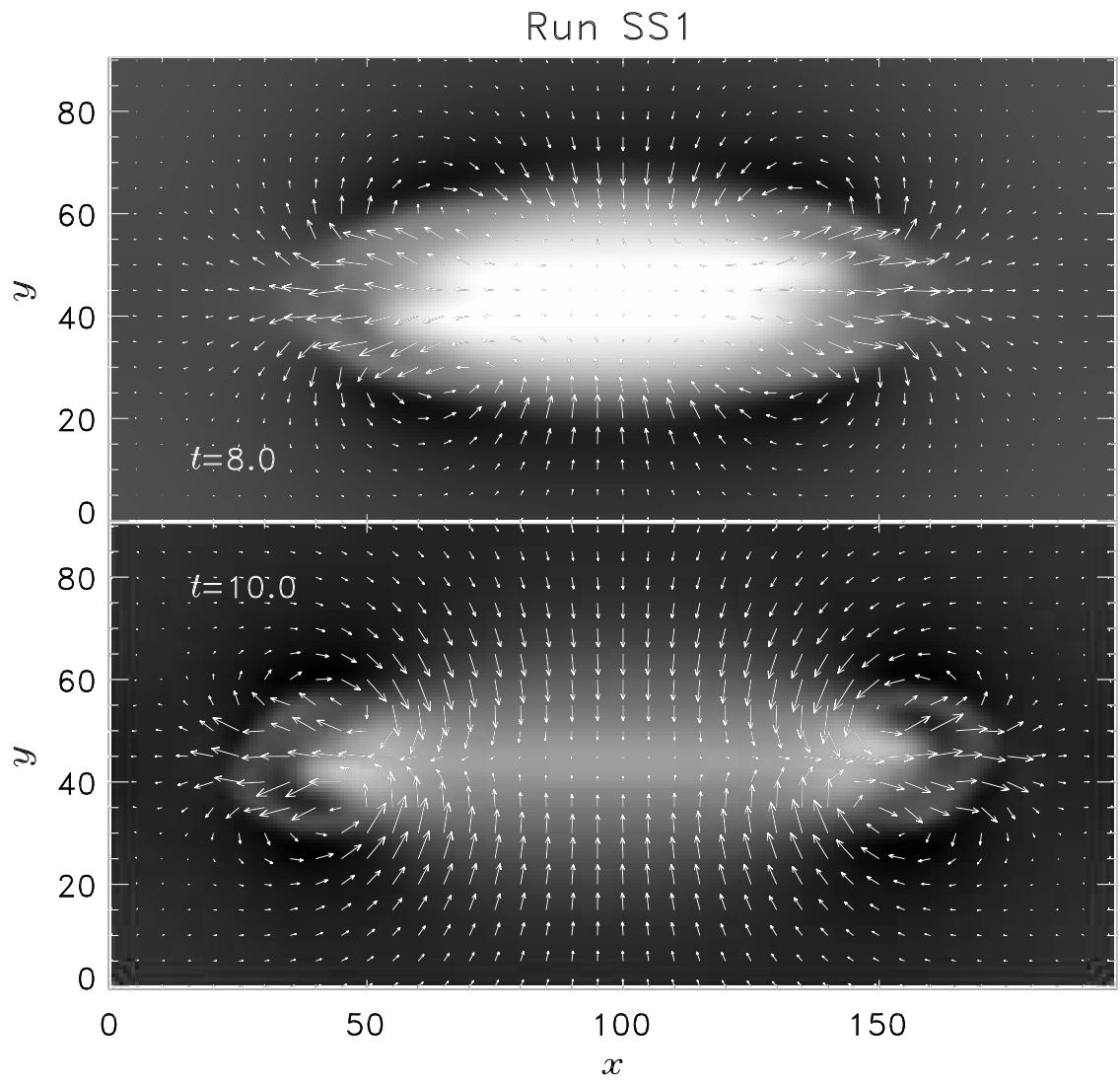,scale=.9,bbllx=75,bbury=550}
\caption[f15.ps]{Same as Figure~\ref{fig14} for the times
corresponding to Figure~\ref{fig13}. \label{fig15}}
\end{figure}

Figures~\ref{fig12} and \ref{fig13} make up a time-series of simulated 
vector magnetograms for the rising, fragmented flux tube of run SS1.  
These images were generated by taking a horizontal slice 
as close to the top of the computational domain as possible without
having the topological evolution affected by interaction with 
the upper boundary.  The grey-scale background represents 
the vertical component of the magnetic field; light regions correspond 
to outwardly directed magnetic field (toward the observer), and 
dark regions correspond to inwardly directed field (away from the observer).  
The arrows define the direction and relative strength of the transverse 
components of the field.  Figures~\ref{fig14} and \ref{fig15} show the 
corresponding velocity field at the same time intervals and 
locations.  The first frame of Figure~\ref{fig12} shows the tip of the 
magnetic sheath as it begins to cut through the horizontal plane.  At 
this point, there is little to no separation between regions of opposite 
polarity, and the corresponding flow field surrounding the magnetic region 
is relatively weak.  This stage is short-lived, since the magnetic
sheath is very thin.  

As the tube continues to rise, the horizontal cut passes through
regions of magnetic field concentrated in the sheath on either side of
the tube apex (which by now has emerged through the 
photosphere), as well as the field concentrated in the relatively
horizontal trailing vortices of each fragment.  As a result, an 
oval-like structure develops, as shown in frame 2 of Figure~\ref{fig12}.
The velocity field has a strong vertical component near the interfaces 
between the vortex tubes and the surrounding non-magnetized plasma, while
exhibiting strong horizontal components at each end. The horizontal
components reflect the flow of material along the sheath away from the 
apex (see frame 2 of Figure~\ref{fig14}).  In frame 1 of Figure~\ref{fig13}, 
the apex of each individual fragment has passed through the cutting plane, 
and we see only the portions of the tube fragments and sheath located 
just above the $\kappa$ inflection point of the $\Omega$-loop.  The 
two regions of opposite polarity now appear crescent-shaped as they 
continue to separate.  After the fragmented portion of the loop has 
emerged through the photosphere, only the more concentrated, 
non-fragmented portions remain.  Thus, in the final frame of 
Figure~\ref{fig13}, the regions of strong magnetic field appear to 
have coalesced --- a feature of emerging active regions that is
commonly observed.

\begin{figure}
\epsfig{file=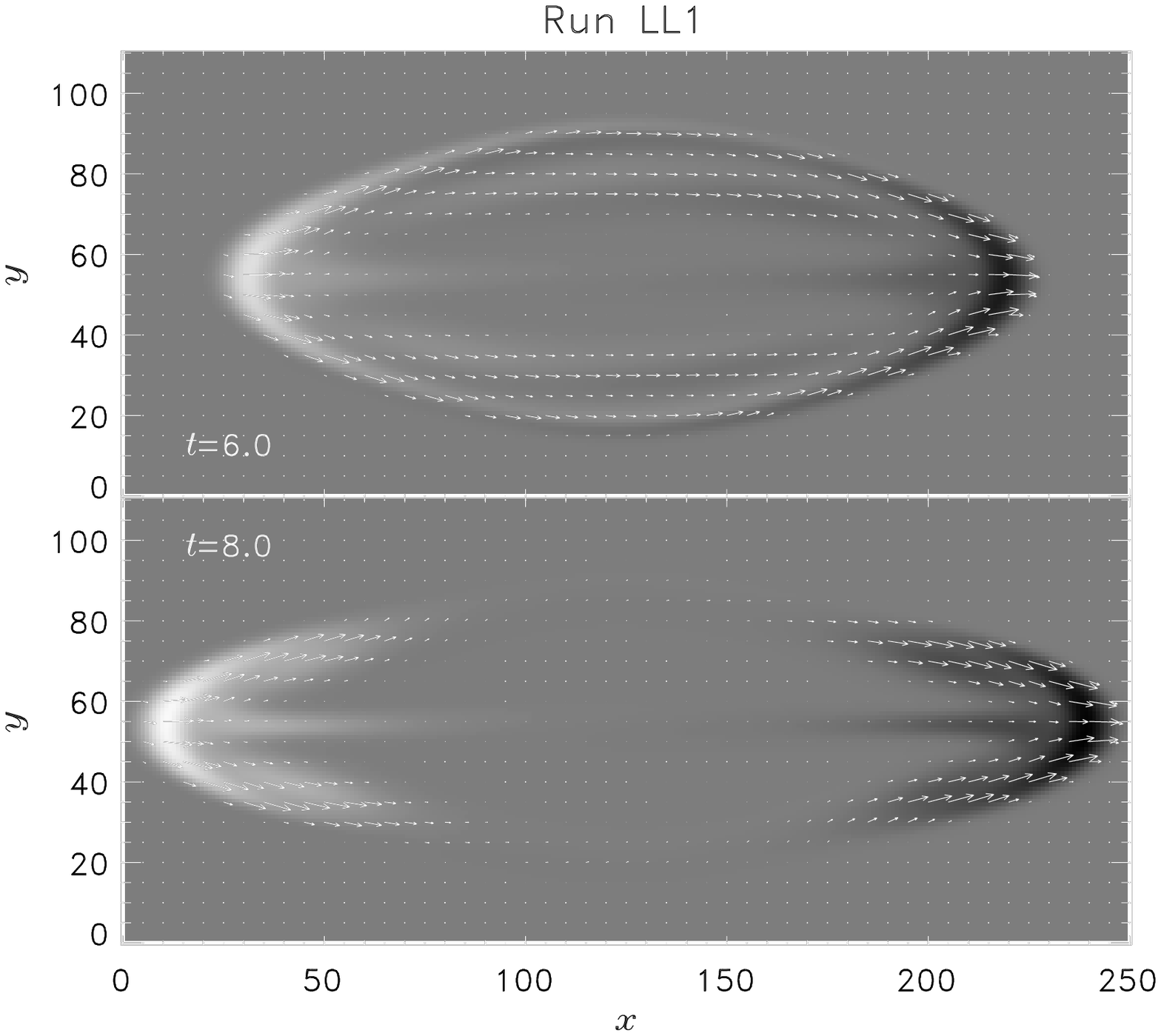,scale=.9,bbllx=75,bbury=550}
\caption[f16.ps]{Vector magnetogram images for run LL1.
The times given are in naturalized units ($t=H_r/[v]$, see text), and
the $x$ and $y$ scale are given in terms of the computational grid.
The grey-scale background represents the vertical component of the
magnetic field (positive values are indicated by the light regions),
and the arrows represent the transverse components.  \label{fig16}}
\end{figure}

\begin{figure}
\epsfig{file=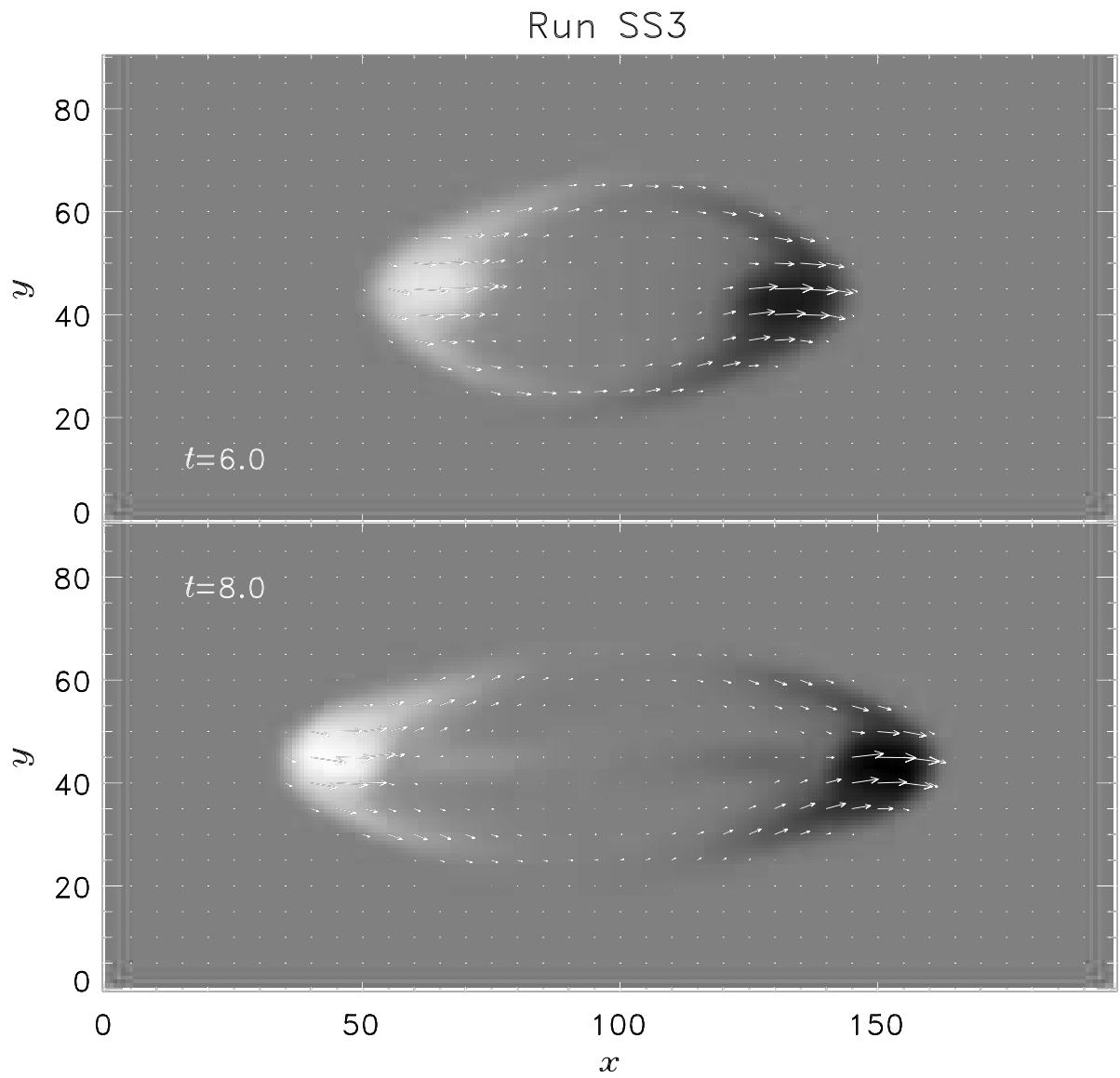,scale=.9,bbllx=75,bbury=550}
\caption[f17.ps]{Same as Figure~\ref{fig16} for run SS3.
\label{fig17}}
\end{figure}

Figures~\ref{fig16} and \ref{fig17} show a similar time-series
of vector magnetograms for runs LL1 and SS3 respectively.  A
comparison between these figures and Figures~\ref{fig12} and \ref{fig13}
reveals the effect of initial twist and tube curvature on the overall 
characteristics of the emerging flux.  A comparison of the surface 
separation of the individual fragments (the semi-minor axes of the 
ellipse-like structures shown in frame 2 of Figure~\ref{fig12} 
and frame 1 of Figure~\ref{fig16}) can be used to determine the relative 
curvature between the two cases.  In general, the lower the apex 
curvature of the fragmented $\Omega$-loop, the higher the ``eccentricity'' 
of the ellipse-like structure.  Similarly, if the crescent-shape of 
the opposite polarity magnetic structures is less distinct or 
absent in the simulations (compare frame 1 of Figure~\ref{fig13} with 
frame 2 of Figure~\ref{fig17}), this indicates that the 
initial level of twist of the magnetic flux tube was relatively high. 
Finally, asymmetries introduced by the higher level of twist
present in the runs where $q=1/4$ can result in the emergence of bipolar
regions that are tilted slightly with respect to the \boldmath $\hat{x}$
\unboldmath direction.  Note that a very slight tilt of approximately 
$4$ degrees from the horizontal has developed between the center of the 
bipolar regions shown in frame 1 of Figure~\ref{fig17}.

Our description of a fragmented flux tube is in qualitative agreement 
with recent images of certain emerging active regions observed by 
\cite{ccf96}, \citet{sz99}, and \citet{ml99}.  Vector magnetograms and 
white-light images of flux emergence taken with high temporal resolution
by the Imaging Vector Magnetograph at the University of Hawaii's
Mees Observatory (IVM) in relatively uncluttered portions of the solar 
disk \citep{ml99} show evolving magnetic structures that are similar 
to the evolving structures described above.  In particular, it is possible 
to infer the presence of oval-shaped magnetic features which evolve into 
crescent-shaped regions of opposite polarity that rapidly separate.  In 
principle, direct measurements of the rate of separation and coalescence 
of the opposite polarity regions of the field may be used to obtain 
information about the structure and evolution of the sub-surface magnetic 
field.  However, to perform this kind of detailed comparison with 
observation will require a more sophisticated set of models.

\section{Summary and Conclusions}\label{conclusions}

We have performed detailed numerical simulations of the rise
of twisted magnetic flux tubes embedded in a non-magnetic, 
stratified plasma using a code which solves the three-dimensional 
MHD equations in the anelastic approximation.  
The evolving magnetic field is described in terms of its volumetric
flux distribution as a magnetic flux tube which may fragment 
into separate, distinct tubes during its rise toward the 
photospheric boundary.  We find that the degree of fragmentation 
of the evolving magnetic flux tube depends not only on the initial ratio 
of the azimuthal to axial components of the field along the tube (as was 
the case in two dimensions), but also on the three-dimensional 
geometry of the tube as it rises through the convection zone.  The 
principal results of our analysis are the following:
\begin{enumerate}
\item If the ratio of azimuthal to axial components of the 
field along a magnetic flux tube exceeds a critical limit, then it
will retain its cohesion and not fragment as it rises through the 
plasma.  The critical limit occurs when the rise speed is
approximately equal to the maximum Alfv\'en speed of the azimuthal
component of the field.  This reinforces the conclusions of 
\citet{em98}, \citet{mie96}, \citet{lfa96}, and \citet{fzl98}.
\item If the initial amount of field line twist is less than
the critical value, and if the magnetic flux tube rises toward the 
photosphere as an $\Omega$-loop, then the degree of apex fragmentation 
depends on the curvature of the loop --- the greater the apex curvature, 
the lesser the degree of fragmentation for a fixed amount of initial 
twist.
\item  In two dimensions, counter-rotating vortices are
effectively infinite in extent and generate long-range flows which
eventually prevent the continued vertical rise of tube fragments.
This artificial geometric constraint is relaxed in three dimensions, 
and although the fragments continue to experience forces due to the 
interaction of the vortex pairs, those forces are only important
along a short section of the tube.  Thus, the fragments are able to 
rise to the surface.  The forces due to vortex interaction depend 
upon the geometry of the $\Omega$-loop --- the greater the apex 
curvature, the lesser their magnitude.  Thus, highly curved
loops exhibit less fragmentation during their rise.
\item Differential circulation between the apex and footpoint of an
$\Omega$-loop leads to the introduction of new magnetic twist
of opposite sign in each leg of a loop fragment. This twist reduces the 
circulation about the apex of each fragment, and further reduces
the forces acting to separate the fragments of the flux tube.
This result is consistent with the predictions of \citet{em98} and
\citet{m97}.
\item  Though these models do not admit to a direct, detailed comparison 
with observations, it is possible to infer certain general observational 
characteristics of emerging magnetic flux if the field configuration 
is that of a fragmented $\Omega$-loop rising through
the solar surface.  If the magnetic field erupts through a relatively quiet 
portion of the Sun, then we expect that one should observe 
concentrations of vertical flux which resemble expanding oval shapes.
This type of feature should quickly evolve into longer-lived crescent-shaped 
regions of opposite polarity that steadily move away from one another.  These 
extended regions should then coalesce once the fragmented portions 
of the $\Omega$-loop have emerged through the solar surface.

\end{enumerate}

\clearpage

\acknowledgments
This work was funded by NSF grants AST 98-19727 and ATM 98-96316, and by
NASA grant NAGS-8468.  The computations described here were partially 
supported by the National Computational Science Alliance and utilized the 
NCSA SGI/CRAY Power Challenge Array.  Further computational support was 
provided by the National Center for Atmospheric Research under grant 
ATM 98-96316, and additional computations were carried out using NCAR's 
CRAY J90 parallel computing facility.  We would like to thank Bob Stein
for allowing us access to additional computational resources, and the 
authors of the FFTW package \citep{fj97} for making their source code 
publicly available.

\clearpage

%% Figures
%
\topmargin=-20pt
\pagestyle{empty}
%
%% Table 

\begin{deluxetable}{cccccccccccc}
\footnotesize
\tablecaption{Initial Flux Tube Parameters \label{tbl-1}}
\tablewidth{0pt}
\tablehead{
\colhead{Label} & \colhead{Resolution\tablenotemark{a}} 
   & \colhead{$L$\tablenotemark{b}} 
   & \colhead{$\mathcal{L}$\tablenotemark{b}} & \colhead{$q$} 
   & \colhead{$Q\!=\!\frac{Lq}{2\pi a}$} 
   & \colhead{$\kappa$\tablenotemark{c}} 
}
\startdata
SS1 & 256$\times$128$\times$128 & 4.362  & 0.0852 & 0.1250 & 0.8678 
  & 1.66248 \\
SS2 & 256$\times$128$\times$128 & 4.362  & 0.0852 & 0.1875 & 1.3017 
  & 1.62260 \\
SS3 & 256$\times$128$\times$128 & 4.362  & 0.0852 & 0.2500 & 1.7356 
  & 1.62906 \\
SL0 & 256$\times$128$\times$128 & 4.362  & 0.1704 & 0.0000 & 0.0000 
  & 1.05930 \\
SL1 & 256$\times$128$\times$128 & 4.362  & 0.1704 & 0.1250 & 0.8678 
  & 1.04928 \\
SL2 & 256$\times$128$\times$128 & 4.362  & 0.1704 & 0.1875 & 1.3017 
  & 0.99512 \\
SL3 & 256$\times$128$\times$128 & 4.362  & 0.1704 & 0.2500 & 1.7356 
  & 1.02579 \\
LS1 & 512$\times$128$\times$128 & 8.724  & 0.1704 & 0.1250 & 1.7356 
  & 0.53913 \\
LS2 & 512$\times$128$\times$128 & 8.724  & 0.1704 & 0.1875 & 2.6034 
  & 0.51325 \\
LS3 & 512$\times$128$\times$128 & 8.724  & 0.1704 & 0.2500 & 3.4712 
  & 0.46269 \\
LL1 & 512$\times$128$\times$128 & 8.724  & 0.3408 & 0.1250 & 1.7356 
  & 0.23734 \\
LL2 & 512$\times$128$\times$128 & 8.724  & 0.3408 & 0.1875 & 2.6034 
  & 0.20638 \\
LL3 & 512$\times$128$\times$128 & 8.724  & 0.3408 & 0.2500 & 3.4712 
  & 0.16699 \\
L1\tablenotemark{d} & 2$\times$128$\times$128 & $\infty$ & $\infty$ 
    & 0.1250 & $\infty$ & 0.00000 \\
L2\tablenotemark{d} & 2$\times$128$\times$128 & $\infty$ & $\infty$ 
    & 0.1875 & $\infty$ & 0.00000 \\
L3\tablenotemark{d} & 2$\times$128$\times$128 & $\infty$ & $\infty$ 
    & 0.2500 & $\infty$ & 0.00000 
\enddata
\tablenotetext{a}{($x$,$y$,$z$)} 
\tablenotetext{b}{in units of $H_r$}
\tablenotetext{c}{apex curvature for last timestep of each run 
   ($\times 10^{-3}$)}
\tablenotetext{d}{two-dimensional limiting case} 
\end{deluxetable}

\end{document}